\documentclass[useAMS,usenatbib]{mn2e}
\usepackage{psfig}
\usepackage{longtable}
\usepackage{rotating}
\usepackage{lscape}
\usepackage{times}
\usepackage{txfonts}

\newcommand{\logns}{$N(>S)~$}

\newcommand{\einstein}{{\it Einstein} }
\newcommand{\chandra}{{\it Chandra} }
\newcommand{\rosat}{{\it ROSAT} }
\newcommand{\sax}{{\it Beppo}SAX }

\newcommand{\xmm}{XMM-{\it Newton} }
\newcommand{\asca}{{\it ASCA }}
\newcommand{\rixos}{{RIXOS} }
\newcommand{\etal}{{\it et~al.$~$}}

\newcommand{\ergs}{{erg~cm$^{-2}$~s$^{-1}~$}}
\newcommand{\ergsnos}{{erg~cm$^{-2}~$}}
\newcommand{\ns}{$N(S)~$}
\newcommand{\hr}{$13^{H}$}

\title[XMM-Newton Deep Field: X-ray source catalogue]
      {XMM-Newton \hr~ Deep field - I.  X-ray sources}
\author[N. S. Loaring \etal]{N. S. Loaring$^{1}$ 
\thanks{nsl@mssl.ucl.ac.uk}, T. Dwelly$^1$, M. J. Page$^1$,  K. Mason$^1$,
I. McHardy$^2$, K. Gunn$^2$, 
\newauthor D. Moss$^2$, N. Seymour$^3$,
A. M. Newsam$^4$, T. Takata$^5$, K. Sekguchi$^5$, T. Sasseen$^6$,
\newauthor F. Cordova$^7$\\
$^1$MSSL, University College London, Dorking, Surrey, RH5 6NT, UK \\ 
$^2$School of Physics and Astronomy, University of Southampton, Southampton,
SO17 1BJ, UK\\
$^3$Institut d'Astrophysique de Paris, 98bis, Boulevard Arago, 75014 Paris,
France\\
$^4$Astrophysics Research Institute, Liverpool John Moores University, Twelve
Quays House, Egerton Wharf, Birkenhead, CH41 1LD, UK\\
$^5$National Astronomical Observatory of Japan, 650 North A'ohoku Place, Hilo,
HI 96729, USA\\
$^6$Department of Physics, University of California, Santa Barbara, CA 93106, USA\\
$^7$University of California, Riverside, 900 University Avenue, Riverside, CA
92521, USA\\}

\begin{document}

\maketitle

\begin{abstract}
We present the results of a deep X-ray survey conducted with \xmm$\!\!$, centred on 
the UK ROSAT \hr~ deep field area. This region covers 0.18 $\rm{deg^{2}}$ and is
the first of two areas covered with
\xmm as part of an extensive multi-wavelength survey designed to study
the nature and evolution of the faint X-ray source population.
We have produced detailed Monte-Carlo simulations to obtain a quantitative 
characterisation of the source detection procedure and to assess the
reliability of the resultant sourcelist. 
We use the simulations to establish a likelihood threshold 
above which we expect less than 7 (3\%) of our sources to be spurious.
We present the final catalogue
of 225 sources. Within the central 9\arcmin, 68 per cent of source 
positions are accurate to 2\arcsec$\,$ making
optical follow-up relatively straightforward.
We construct the \logns relation in four energy bands: 0.2-0.5 keV, 0.5-2 keV,
2-5 keV and 5-10 keV. 
In all but our highest energy band we find that the source counts can be
represented by a double powerlaw with a bright end slope consistent with the
Euclidean case and a break around $10^{-14}$ \ergs$\!\!$.
Below this flux the counts exhibit a flattening. 
Our source counts reach densities of 700, 1300, 900 and 300 deg$^{-2}$ 
at fluxes of
$4.1\times10^{-16}$, $4.5\times10^{-16}$, $1.1\times10^{-15}$ and
$5.3\times10^{-15}$ \ergs in the 0.2-0.5, 0.5-2, 2-5 and 5-10 keV energy
bands respectively. We have compared our source counts with those in the 
two \chandra deep fields and Lockman hole and find our source counts to be  
amongst the highest of these fields in all energy bands. 
We resolve $>51$\% ($>50$\%) of
the X-ray background emission in the 1-2 keV (2-5 keV) energy bands. 

\end{abstract}


\section{Introduction}
It is widely accepted that the majority
of the Cosmic X-ray Background (XRB) arises from the integrated emission of
discrete extragalactic sources \citep{schwartz76,giacconi87,maccacaro91}.
The energy density of the XRB peaks at $\sim$30 keV, but the first imaging
surveys were carried out at much lower energies: $<$3.5 keV with \einstein and
$<$2 keV with \rosat. By the late 1990s \rosat surveys had resolved 
70-80\% of the soft XRB, \citep{shanks91,hasinger93, hasinger98, mchardy98}. 
Subsequently, deep \xmm and \chandra surveys have essentially
resolved the soft XRB into discrete sources \citep{mushotzky00,hasinger01,
brandt01,tozzi01,rosati02,alexander03}. 
Optical follow up of these sources has revealed a population
composed primarily of unobscured broad line active galactic nuclei (AGN),  with
an increasing fraction of absorbed AGN at fainter fluxes
\citep{mchardy98,schmidt98,zamorani99,lehmann01,szokoly03,barger03}.

In order to investigate further the nature of the obscured population one has
to conduct surveys at harder energies ($>$2 keV), which are less
sensitive to absorption. 
Surveys carried out using \asca
\citep{georgantopoulos97,cagnoni98,ueda98,ueda01, ishisaki01} and \sax
\citep{fiore99,giommi00,fiore01} resolved 25-35\% of the 
XRB above 2 keV.  
More recently, the  \xmm and \chandra deep field surveys have resolved 60-90\%
of the hard ($>$2 keV) X-ray background \citep{hasinger01,
giacconi01,tozzi01,cowie02,rosati02,alexander03,manners03} 
probing fluxes a factor 100$\times$
fainter than the \asca and \sax surveys. The wide range in resolved fraction
arises not only due to the variation in source counts between the 
surveys, but also due to the uncertainty of the absolute normalisation of the
hard XRB, for example the \sax XRB normalisation from 
\citet{beppoxrb} is $\sim$30\% higher than the ASCA value from
\citet{gendreau95}. 

Optical follow up studies of the deepest surveys
find a predominance of ${z}<1$ objects which do not show broad
emission lines \citep{barger01,tozzi01,barger02,lamer03,barger03}. 
There is an increasing contribution from normal galaxies at the faintest 
fluxes, and it appears likely that normal galaxies will outnumber AGN 
below 0.5-2 keV fluxes 
of $10^{-17}$ \ergs \citep{bauer04}. A significant fraction
of the hard X-ray sources in these fields are optically faint, with $R>24$
and are therefore difficult to identify optically \citep{alexander01,mchardy03,
alexander03}.

Despite the great advances made in detecting increasingly fainter X-ray
sources, the physical nature and evolution of the faint hard X-ray population 
remains largely unknown. The redshift distribution and column density
distribution of the absorbed AGN are still poorly constrained.

Another issue which needs addressing is the relationship between gas and 
dust absorption in AGN. There
have been several cases of a mismatch between optical and X-ray classifications,
indicating a wide range in dust/gas ratios for obscured sources
\citep{akiyama00,page01,comastri01,maiolino01a,loaring03,carrera04}.

In particular, high quality X-ray spectra are needed 
to determine the dominant X-ray emission mechanisms and the
amount of absorption. We have therefore used \xmm to carry out deep surveys of
two widely separated fields to probe the X-ray
population down to fluxes $\sim10^{-15}$ \ergs in the 0.5-2 and 2-5 keV energy
bands. The source counts in these energy bands exhibit a
break at $\sim10^{-14}$ \ergs \citep{rosati02} around which 
the maximum contribution to the XRB per logarithmic flux
interval occurs. 

This paper presents the X-ray catalogue derived from the first of these two \xmm surveys,
carried out in the UK \rosat deep field area (hereafter the \hr~ deep field).  
The field has also been observed with a  mosaic of \chandra 
pointings which cover the whole \xmm field of view and provide accurate
source positions. It is complemented with multiwavelength follow up in the UV,
optical, near-IR, mid-IR and radio \citep{seymour04}. The \chandra catalogue has already been
presented elsewhere \citep{mchardy03}; here we present the \xmm catalogue and
observed source counts. 

\section{Observations and data reduction}
\label{obs}

\subsection{XMM observations}
The \hr~ deep field is centred on the sky co-ordinates RA 13h 34m
37.1s, Dec +37$^\circ$ 53$'$ 02.2\arcsec~ (J2000).  The XMM-Newton
observations were carried out in three separate revolutions during June 2001
for a total exposure time of 200 ks.  The
European Photon Imaging Cameras \citep[EPICs][]{turner2001} 
were operated in standard full-frame mode. 
The thin filter was
used for the PN camera, while the thin and medium filters were
alternated for the MOS1 and MOS2 cameras. Table
\ref{tab:observing_log} gives a summary of the observations.

The data were processed using the \xmm Standard Analysis System (SAS)
version 6.0.  Approximately 40\% of the total observation time was
affected by high particle background flares, arising from soft protons hitting the detector. 
The data were therefore temporally filtered to remove these high background
periods. In practice, times where the 5-10 keV count rates exceeded 2 s$^{-1}$ in the MOS
cameras and 4 s$^{-1}$ in the PN camera were excluded. Filtering
reduced the total useful exposure time from $\sim$200 ks to $\sim$120 ks.
The net live times for the individual detectors after the periods of
high background were excised are listed in Table
\ref{tab:observing_log}.

A significant component of the EPIC background comes from instrumental 
emission lines, in particular the Cu K$\alpha$ line at 8.1 keV 
in the PN 
and the Al K$\alpha$ line at 1.5 keV in
both  detectors \citep{lumb02}.
Events with energies close to those of 
the emission lines were filtered out to minimise the instrumental 
contribution to the background. 
Events in bad columns, bad pixels and adjacent to chip edges were also 
excluded.

Images and exposure maps were then constructed from each observation for each
detector in four energy bands: 0.2-0.5 keV, 0.5-2 keV, 2-5 keV
and 5-10 keV. Single-pixel events were used to construct 
the PN 0.2-0.5 keV image. Single, double and triple events were used to 
construct the higher energy PN images. For MOS, all valid event patterns were 
used to construct the images regardless of energy band.
In each energy band, the exposure maps were scaled to
the PN thin filter response. The images and exposure maps from the different 
detectors and observations were then summed to produce one
image and one exposure map per energy band. 
The response-weighted summation over each observation and telescope
gives total on-axis PN-equivalent live exposure times of 152 ks, 161 ks,
179 ks, and 160 ks, in the 0.2-0.5 keV, 0.5-2 keV, 2-5 keV,
and 5-10 keV bands respectively.

For the EPIC imaging observing modes, photons are not only registered during
the actual integration interval but also during the readout of the CCD. These
out-of-time events are hence assigned the wrong position in the readout
direction. The fraction of out-of-time events is highest for the PN full
frame mode (6.3 \%) and therefore for each PN exposure, an additional
synthetic out-of-time events list was produced by randomising the coordinates
of the events within each chip  in the readout direction. Out-of-time images 
were constructed in each energy band by filtering these event lists in exactly
the same way as the real event lists. These out-of-time images were used as
inputs to the background model as described in Section \ref{sec:background}.

The astrometry of the individual observations was corrected for small offsets
between the pointings. A sourcelist was constructed for each observation as
described in Section \ref{sec:background} and cross-correlated with the optical
positions of the 214 sources found in the \chandra catalogue of \citet{mchardy03}
using the SAS task EPOSCOR. The appropriate offsets in RA and dec were then
applied to each of the individual events to tie the \xmm data to the 
optical/\chandra$\!$/radio co-ordinate frame. The images and exposure maps were then
reproduced with the correct astrometry. 
The actual offsets in RA and Dec differed slightly between the three
observations. The first observation (revolution 276, 179 \xmm sources) had offsets of 1.4\arcsec,
-1.3\arcsec~ applied, the second (revolution 281, 106 \xmm sources) had offsets of 0.5\arcsec,
-0.5\arcsec~ applied and the third observation (revolution 282, 257 \xmm sources) had
offsets of 0.6\arcsec, 0.2\arcsec applied.

\begin{table}
\begin{center}
\scriptsize
\begin{tabular}{ccccccc}
\hline \hline

 Rev. & Date &\multicolumn{3}{c}{Live Time (ks)}&\multicolumn{2}{c}{Filter} \\
      &      & MOS1 & MOS2 & PN                 & MOS1 & MOS2 \\
\hline
 276 & 12.06.01 & 43.1 & 45.8 & 35.5 & thin & med  \\
 281 & 22.06.01 & 14.1 & 12.0 & 6.7 & med  & thin \\
 282 & 23.06.01 & 59.2 & 60.2 & 47.9 & thin & thin \\

\hline

\end{tabular}
\normalsize
\caption{Summary of \hr~ deep field \xmm observations showing the date and
length of observations and the filters used. The live times have
had periods of high background excluded.}
\label{tab:observing_log}
\end{center}
\end{table}

\subsection{\xmm Source detection}
\label{sec:background}
The combined images in each energy band were source-searched
simultaneously using the SAS tasks EBOXDETECT and EMLDETECT.
EBOXDETECT is a sliding cell detection algorithm which outputs an initial
sourcelist. This sourcelist is input for the
EMLDETECT task which  performs a maximum likelihood PSF fit to the sources 
producing refined positions and fluxes for all bands simultaneously. 
This method results in better source positions than searching in the individual
energy bands one at a time because the PSF fit
is based on the maximum number of counts per source. 
If the best fit source count rate in any particular energy band is less than
zero (i.e. there are fewer counts at the source position in the image than in the 
background map) the source count rate is set to zero in this energy band.

A background map was produced for each combination of observation, detector and
energy band (36 background maps in total) using our own software. 
This software performs a maximum likelihood fit to the background, 
assuming a three-component background model: out-of-time events, a flat
unvignetted component, caused primarily by cosmic rays, and a
vignetted component representing unresolved faint sources and
genuinely diffuse emission. For the PN, the out-of-time events
contribution to the background was fixed at 6.3\% of the intensity of
the synthetic out-of-time images; for the MOS background, we assumed
no contribution from out-of-time events. The intensities of the
vignetted and unvignetted background components were free parameters
in the fit.

To maximise sensitivity an iterative procedure was employed.
Initially, sources were detected in each individual image (per detector and per
energy band) using a 3-pixel-square sliding cell
in EBOXDETECT, with the background computed as the average of the
surrounding $7 \times 7$ pixels.  Then, the sources were excised from
the individual images and the background fitted. Each of the background maps 
from the MOS and PN cameras for a given energy band were then summed to produce
one combined background map for each energy band. The resultant background maps
were then used for the sliding cell (EBOXDETECT) followed by maximum likelihood
(EMLDETECT) source detection on the combined image in each energy band. 
The sequence of background determination 
followed by source searching was then repeated several times. We found that 
the sourcelist and background maps converged after 4 iterations. 
Likelihood thresholds (DET\_ML values output from the source detection)
of 4 and 5 were chosen for EBOXDETECT and EMLDETECT
respectively. These values are related
to the probability of a random Poissonian fluctuation having caused the
observed source counts via \citep{cash79}:

\begin{equation}
\rm {DET\_ML = -ln P_{random} } 
\label{eqn:cash}
\end{equation}

The conversion factors from count rates to flux were determined from the EPIC 
response matrices, over exactly the same energy ranges as those in which the 
images were constructed, assuming a power law spectrum with photon index $\Gamma=1.7$.
This is a good average within our flux range \citep{page03,mateos05}; 
however the sources have a range of photon indices. To assess the impact
of such a spread we have calculated the expected conversion factors using
photon indices of $\Gamma=1.4$ and $\Gamma=2.0$ respectively. 
This range represents the expected spread in spectral slope for sources 
contributing to the XRB. 
The relatively flat $\Gamma=1.4$ lower limit corresponds to the XRB slope in
the 3-15 keV range, produced by absorbed AGN. The upper limit value is
typically found in unabsorbed AGN, and we would therefore expect the bulk of
our sources to lie between the two values. 
 
The largest effect is in the
0.2-0.5 keV and 5-10 keV energy bands where the conversion factors derived are
different from those assuming a photon index of $\Gamma=1.7$ by up to 11\% and
8\% respectively. However, in the 0.5-2 and 2-5 keV bands the photon index
chosen only affects the  conversion factors  by 1-2\%. 
The PN response matrices include only single and double pixel events, but 
for the three highest energy bands our PN images also include triple and 
quadruple 
events. In order to take this into account, the count rate to flux 
conversion factors were corrected (by up to 6\% in the hardest band) 
as described in \citet{osborne01}. 

\begin{table}
\begin{center}
\begin{tabular}{lc}
\hline \hline
Energy (keV) & ECF ( cts  per $10^{-11}$ \ergsnos ) \\
\hline
0.2--0.5 & 4.7775\\
0.5--2.0  & 4.8905\\
2.0--5.0  & 1.9605\\
5.0--10.0 & 0.5929\\
\hline
\end{tabular}
\caption{Energy conversion factors (ECF) used to convert between count rate and
flux.}
\end{center}
\label{tab:conversions}
\end{table}

\begin{figure*}
\setlength{\unitlength}{1in}
\begin{picture}(7,3.5)
\put(-0.15,0.1){\includegraphics{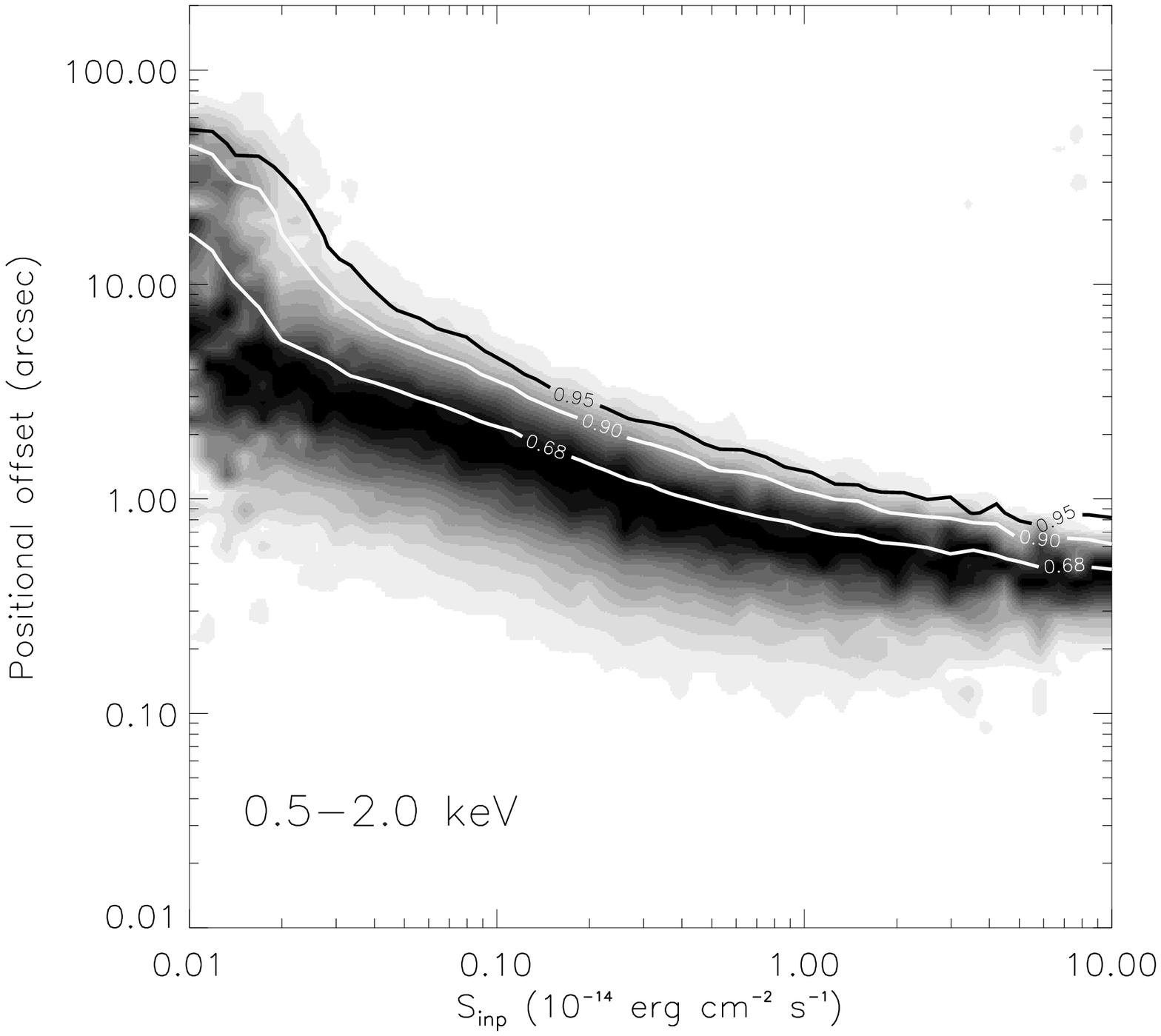}}
\put(3.2,0.1){\includegraphics{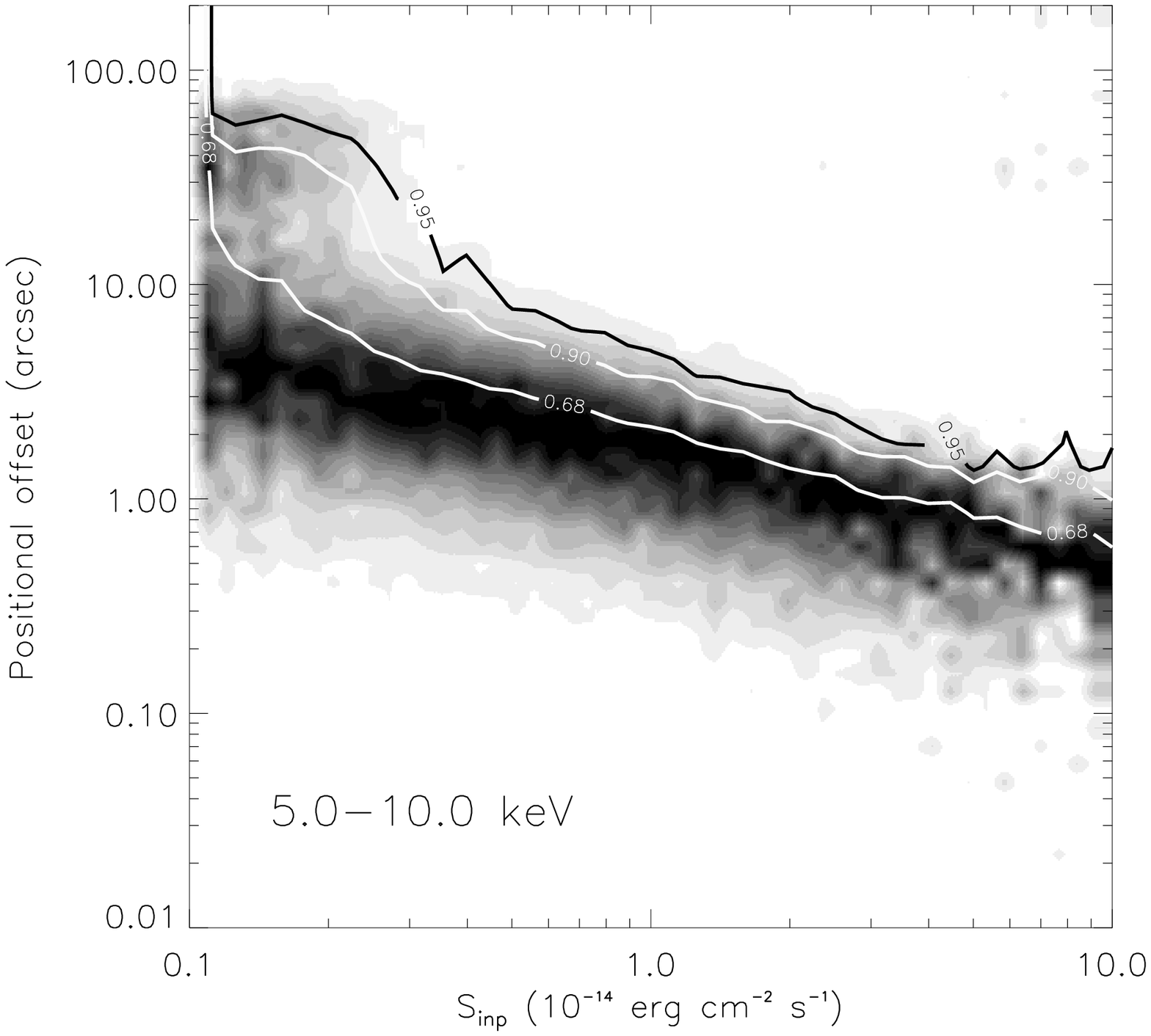}}
\end{picture}
\vspace{-1.0cm}
\caption{Greyscale images showing the distribution of positional offsets
between output and input source locations as a function of input flux in two
energy bands. 
Those sources within the central
9\arcmin~ of the \xmm field of view having $ \rm {DET\_ML \geq
5 }$ are shown. The concentration of positional offsets as a function of
$S_{inp}$ is indicated by the darkness of the greyscale image.
The three contours show the distances within which
68, 90, and 95 per of the data lie.}
\label{fig:flux_posdiff}
\end{figure*}

\section{Monte carlo field simulation}
\label{sec:sims}
We have used an \xmm specific extension of the simulation method of
\citet{hasinger98} to obtain a quantitative characterisation of the source
detection procedure and to assess the reliability of the resultant 
sourcelist. We have used our simulations to find the appropriate detection 
threshold to be applied to the \hr~ field. 
A Monte-Carlo approach is particularly powerful near the survey flux limit where
a number of different processes contribute to uncertainties in the detected
source parameters.
Our simulation method consists of several modular steps that are repeated for a
large number of synthetic fields. 
Briefly, an `input' sourcelist was generated independently in each energy band. 
Each list was then folded 
through the \xmm imaging characteristics to generate images in each energy
band. These images were then source searched to produce  `output'
sourcelists. A pairing algorithm was used to associate an `input'
source with each `output' source. Each of these stages are described in detail
in Appendix A. Here we present the results of a comparison between the output
and input source properties. These provide an indication of the biases inherent
in our survey.
One thousand simulated fields were used to reduce
statistical uncertainties in the analysis. 

\subsection{Comparison of simulated input and output sources}
\label{sec:simcomp}

In order to assess the accuracy of source positions and fluxes, and to
estimate the degree to which confusion and Eddington bias affects the source
counts in our \hr~ data we have compared the input and output properties of our
simulations. The simulations should mimic any biases found in the real data. 

We matched each output source found in the
simulated images to the closest valid input source. We consider an input
source as valid when its flux, $S_{inp}$, contributes a reasonable fraction 
($>$20 per cent) to the total output flux, $S_{out}$. 
No upper limit was applied to the radius at which
input and output sources were matched in order to assess the typical offsets 
between input and output positions.
Fig. \ref{fig:flux_posdiff} shows the distribution of measured positional
offsets as a function of input flux, $S_{inp}$. 
The greyscale image shows the relative density of sources at a given $S_{inp}$.
The dark band shows where the majority of sources
lie.
All sources with DETML values $>5$ and offaxis angles $<9$\arcmin are shown.
The contour lines plotted correspond to the positional offsets within which
68, 90 and 95  per cent of the data lie.

The mean positional offset decreases with increasing flux and is $<10$\arcsec~ for
all but the faintest fluxes. In the 0.5-2 keV energy band, 95 per cent of
sources with $S_{inp}>5\times10^{-16}$ \ergs and offaxis angles $\leq 9$\arcmin~ have
positional offsets $<10$\arcsec. In the 5-10 keV energy band, 95 per cent of
sources with 
$S_{inp}>5\times10^{-15}$ \ergs and offaxis angles $\leq 9$\arcmin~ have
positional offsets $<10$\arcsec.
Those with offaxis angles $>9$\arcmin~ have
systematically larger positional offsets (larger by $\sim2$\arcsec over the
majority of flux ranges).  
Any sources with higher positional offsets are most likely due to incorrect 
associations. A discussion of the positional accuracy found in the real
\hr~ data and its comparison with the simulations is deferred to Section
\ref{sec:comparison}.

Subsequently, we matched output sources to input sources within a radius
$r_{cut}=5\arcsec$, $8\arcsec$, and $10\arcsec$ for offaxis angles of $0-9'$,
$9-12'$ and $>12'$ respectively, reflecting the degradation of the XMM PSF
\citep{kirch04} away from the optical axis. 
Where more than one candidate input source lay within
$r_{cut}$, the brightest was chosen. The 
brightest input candidate within $r_{cut}$ must be the `correct' input
counterpart in the sense that it is the largest contributor to the output 
source counts. 
In practice, when averaged over the 1000
simulations, only 0.6, 1.7, 1.5 and 0.5 output sources per field had more than
one valid ($5S_{inp}\geq S_{out}$) input candidate within $r_{cut}$ in the 
0.2-0.5, 0.5-2.0, 2.0-5.0, and 5.0-10 keV energy bands.

\begin{figure*}
\setlength{\unitlength}{1in}
\begin{picture}(7,3.5)
\put(0.0,-2.0){\includegraphics{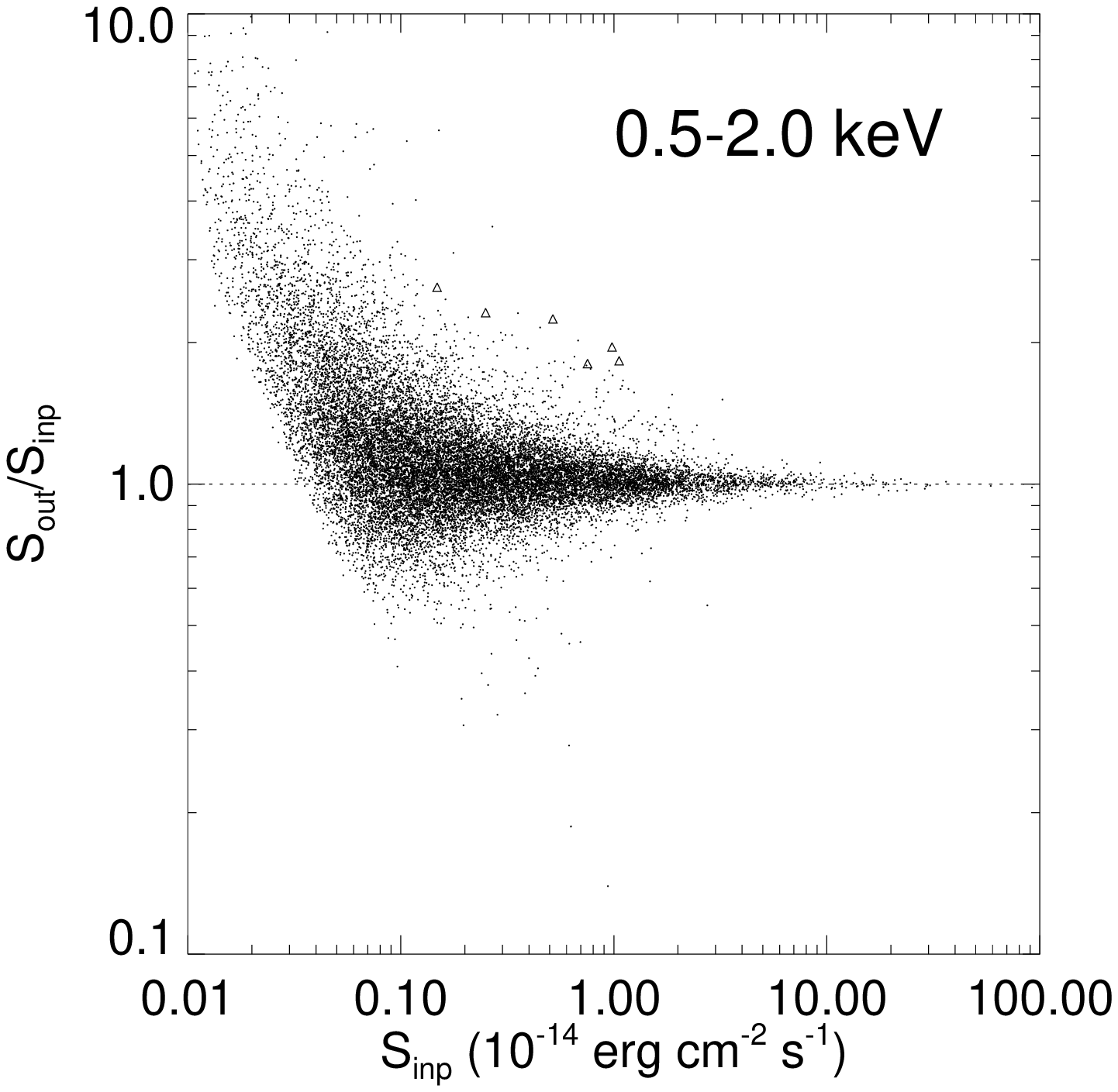}}
\put(3.2,-2.0){\includegraphics{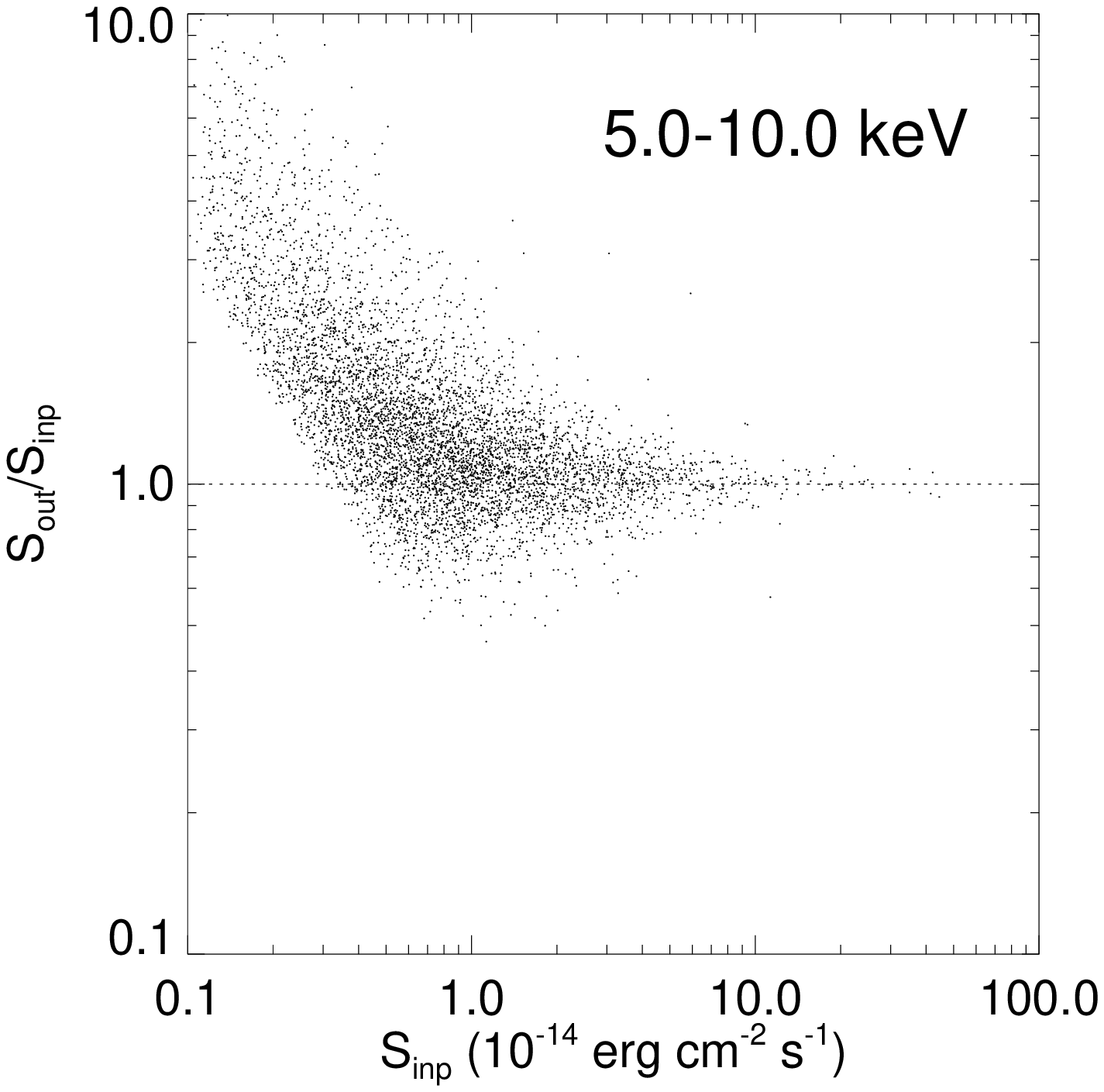}}
\end{picture}
\vspace{-1.3cm}
\caption{Scatter plots of input vs output flux 
in two selected energy bands. We only show those
detected sources having $\rm{DET\_ML \ge 5}$ and input counterparts within
5\arcsec, 8\arcsec, and 10\arcsec~ for input offaxis angles of $0-9'$, $9-12'$
and $>12'$ respectively. Sources with $S_{out}/(S_{inp} + 3\sigma_{out}) >
1.5$ are plotted as triangles. The results for 100 simulations are shown for
clarity.}
\label{fig:flux_in_out}
\end{figure*}
 
\begin{figure*}
\setlength{\unitlength}{1in}
\begin{picture}(6,4.5)
\put(0.1,-4.0){\includegraphics{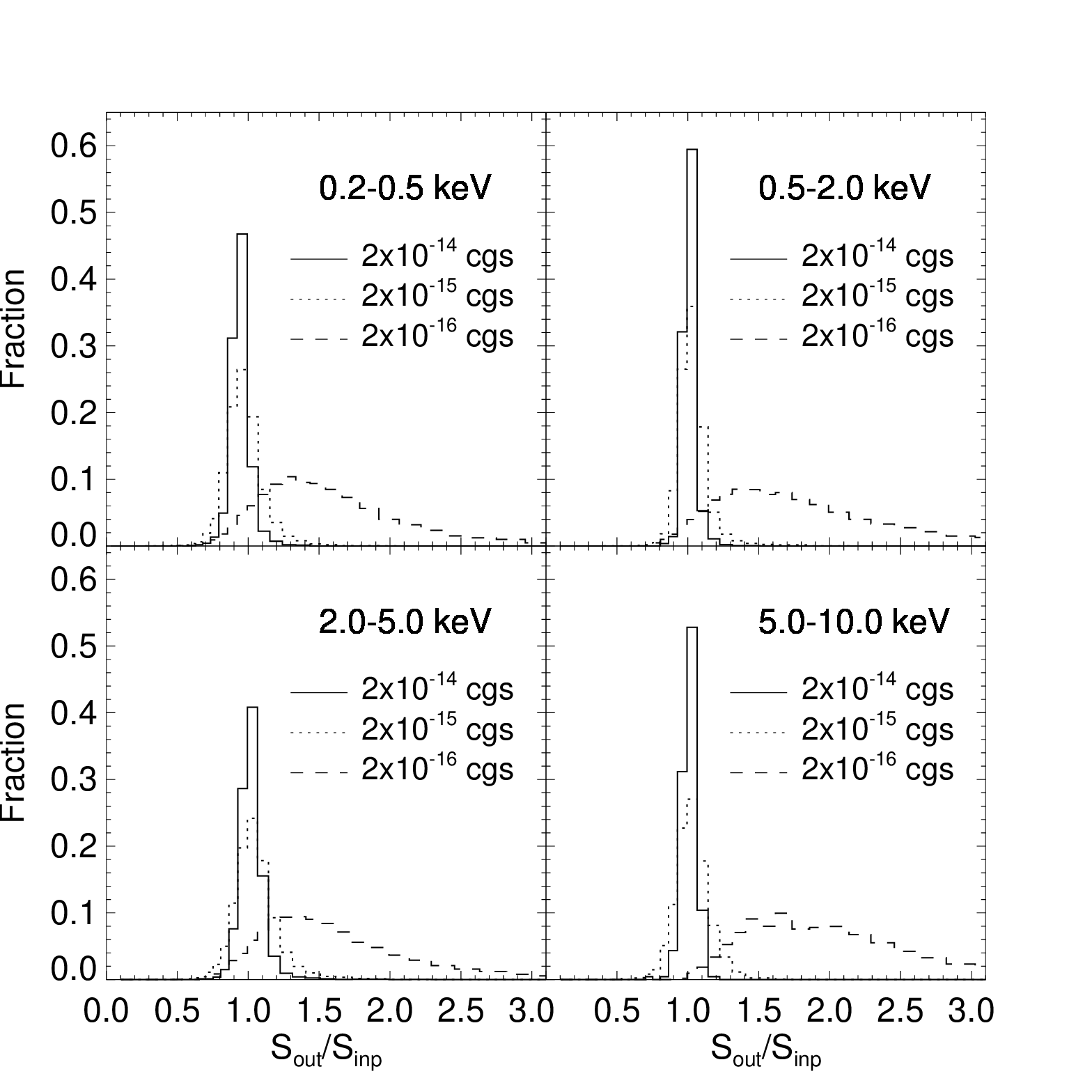}}
\end{picture}
\vspace{0.7cm}
\caption{The distribution in output ($S_{out}$) flux to input flux ($S_{inp}$)
ratio for three illustrative input flux ranges.
The solid line represents the sources in the flux interval centred on
$S_{inp}=2\times10^{-14}$ \ergs ($1\times10^{-14}-4\times10^{-14}$ \ergs). 
The dotted line represents sources in the flux
interval centred on $S_{inp}=6\times10^{-15}$ \ergs ($3\times10^{-15} -
1.2\times10^{-14}$ \ergs) and the dashed
line represents sources in the flux interval centred on $S_{inp}=4\times10^{-16}$
\ergs ($2\times10^{-16} - 8\times10^{-16}$). The flux intervals were chosen 
such that $\Delta$log$S_{inp}=0.3$.}
\label{fig:hist}
\end{figure*}

\begin{figure*}
\setlength{\unitlength}{1in}
\begin{picture}(7,3.5)
\put(-0.1,-2.0){\includegraphics{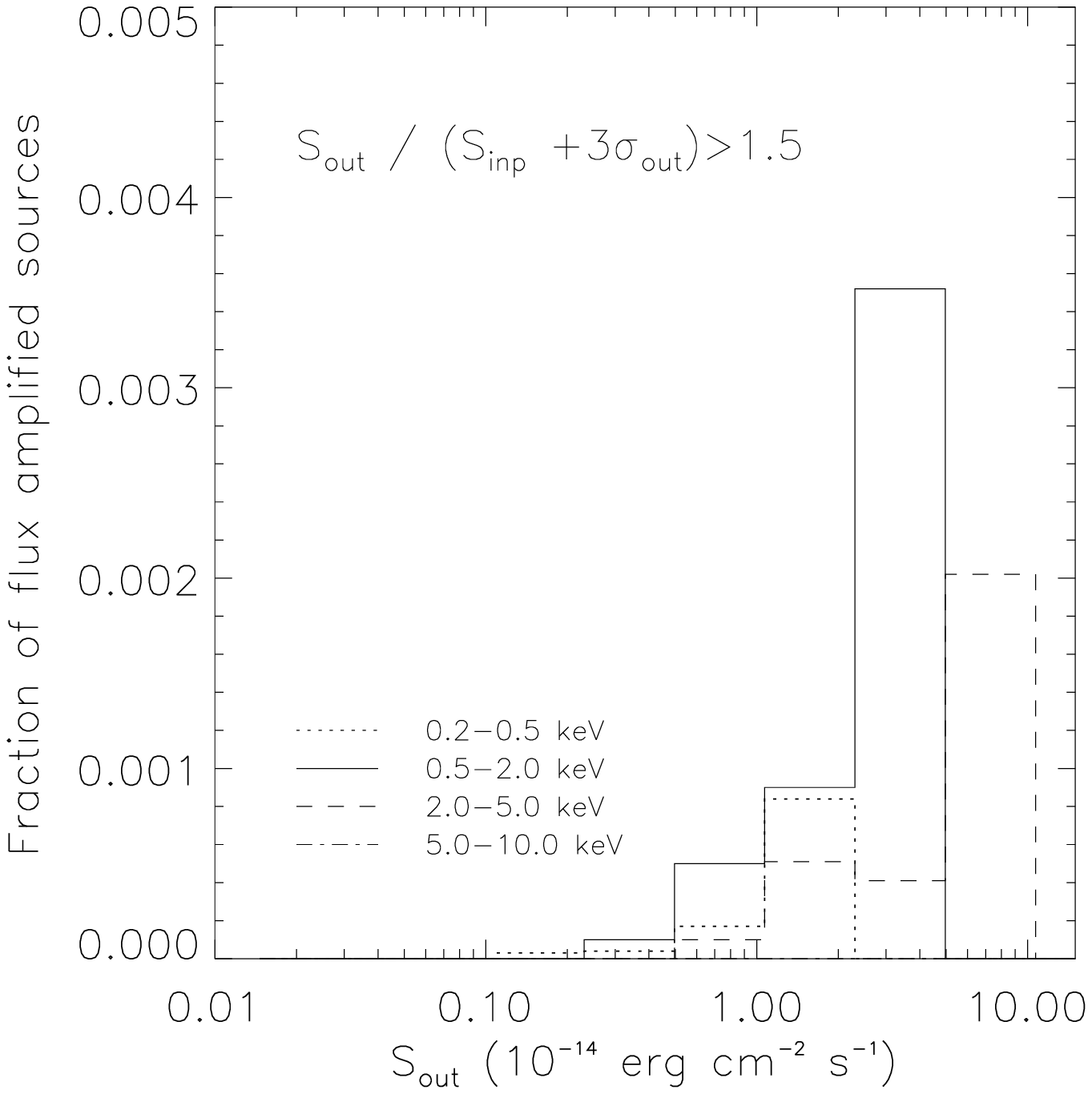}}
\put(2.75,2.9){{\Large{a}}}
\put(3.1, -2.0){\includegraphics{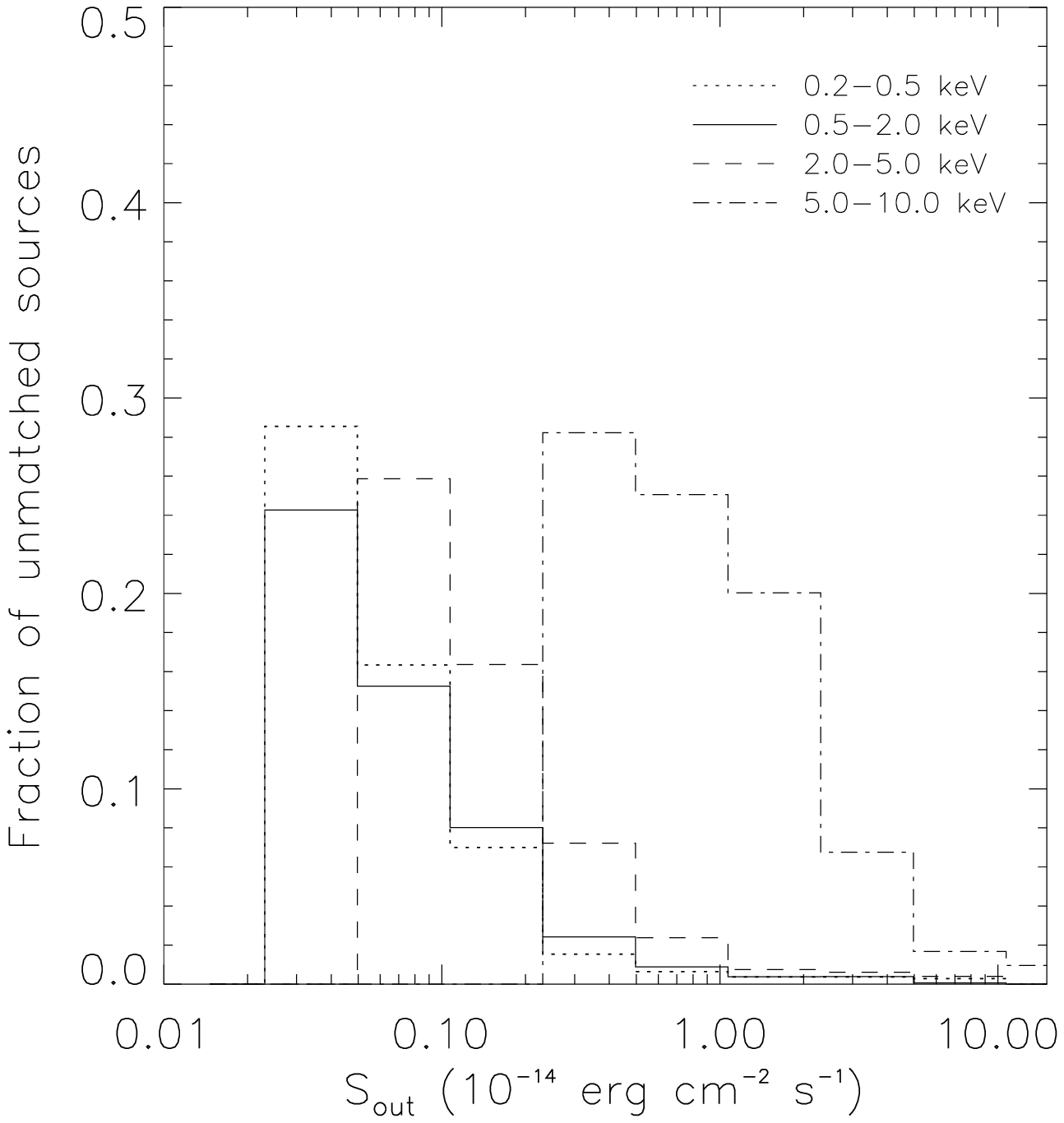}}
\put(4.35,2.9){{\Large{b}}}

\end{picture}
\vspace{-1.0cm}
\caption{a) Fraction of flux amplified sources as a function of flux. b) Fraction of
unmatched sources as a function of flux. (0.2-0.5 keV (dotted), 0.5-2 keV
(solid), 2-5 keV (dashed), 5-10 keV (dot-dashed)). The fraction of flux
amplified sources is $<0.4$ per cent in all energy bands and is significantly less
than the fraction of unmatched sources.
}
\label{fig:confusion}
\end{figure*}

The output fluxes and positions were then compared with the corresponding
input values. 
There are several reasons why we might expect a difference between the input
and output flux distributions and source counts: i)
a systematic or statistical flux measurement inaccuracy, 
ii) source confusion, iii) statistical fluctuations in the background which
may be detected as sources, and iv) Eddington bias. 
All these factors must be considered together when 
interpreting the \hr~ field data.

The intrinsic accuracy of the source detection photometry
is best evaluated at high fluxes and low offaxis angles, 
where ii), iii) and iv) are less important.
The bright end of Fig. \ref{fig:flux_in_out} illustrates the high fidelity of
the detected fluxes, $S_{out}$, to the input fluxes, $S_{inp}$. 
Considering sources with
$S_{inp}>5\times 10^{-14}$ \ergs$\!$, with offaxis angles $<9'$ the average 
$S_{out}/S_{inp}$ ratio is 1.01$\pm0.03$ (1$\sigma$) in energy bands 0.2-0.5,
0.5-2 and 2-5 keV. In our highest energy band the average $S_{out}/S_{inp}$
ratio is also 1.01 but the scatter is larger ($1\sigma=0.06$).
   
The distribution of $S_{out}/S_{inp}$ is shown in
Fig. \ref{fig:hist} for three flux intervals in each energy band. 
At bright fluxes ($S_{inp}=2\times10^{-14}$ \ergs) the distribution is narrow,
symmetrical and centred on
$S_{out}/S_{inp}=1$, because the statistical errors on the fluxes are small. 
At intermediate fluxes ($S_{inp}=6\times10^{-15}$ \ergs) the distributions are
still
relatively symmetrical, but they are slightly broader as the statistical errors
on the fluxes are larger. However, at the faintest fluxes, the distributions
are much broader and significantly skewed towards larger $S_{out}/S_{inp}$
ratios. The increased width is due to the increased statistical errors on the
fluxes. The distributions are shifted towards larger $S_{out}/S_{inp}$ ratios
because at such faint fluxes sources are unlikely to be detected unless they
are enhanced by Poisson fluctuations or by source confusion.

Source confusion occurs when two or more nearby input sources fall in a single
resolution element of  the
detector and result in a single output source. This results in a flux
amplification in the output source and a net loss of fainter sources. 
The position of the output source will be close to the centroid of the 
merged input sources. Therefore when two input sources of similar flux are
confused, the output position does not correspond to either of the input
positions.
Source confusion can limit the depth of any deep survey
depending on the size of the telescope beam, and on the sky-density of objects
as a function of flux. 
In practice, we cannot distinguish between 
a source boosted by photon noise or one confused with another faint source,
therefore we consider the two effects jointly.
We class sources as `flux amplified' (corresponding to `confused'
sources in \citealt{hasinger98}) if
$S_{out}/(S_{inp}  + 3\sigma_{out}) > 1.5$ (where $\sigma_{out}$ is the
1$\sigma$ error on the output flux $S_{out}$).
Fig. \ref{fig:confusion}a shows
the fraction of flux amplified sources as a function of input flux in our four
energy bands.
The fraction is less than 0.4 per cent at all fluxes in all energy bands.
This fraction will depend on the exact definition of flux 
amplification. 
Using a less stringent definition of $S_{out}/(S_{inp} + 3\sigma_{out}) > 1.2$ 
still results in a fraction well below 2\% in each
energy band at all fluxes. 


We class an output source as `unmatched' when there are no valid ($S_{out} \leq
5S_{inp}$) input sources within $r_{cut}$ 
(corresponding to `spurious' sources in \citealt{hasinger98}). 
These are mainly caused by positive fluctuations in the
background. Fig. \ref{fig:confusion}b shows the fraction
of unmatched sources as a function of flux in each energy band. 
As expected, the unmatched fraction is highest at very faint fluxes,
where up to 30\% of the  sources are
unmatched. Unmatched sources are many times more numerous than the flux
amplified sources at any flux. 

In our simulations we curtailed our input sourcelists at fluxes 5$\times$
fainter than those found in the \hr~ data in each energy band in order to speed
up processing time. In order to assess the impact of the simulation flux limit
on the number of flux amplified and spurious sources, we have also produced a
smaller number of simulations to a greater depth, reaching fluxes 10$\times$ fainter
than those found in the \hr~ data.
The fraction of flux amplified sources in these faint simulations agrees with
the fraction found in our original simulations to within 0.02 per cent. 
Likewise, the fraction of unmatched sources agrees to within 2 per cent.
We are therefore satisfied that our chosen flux limits are sufficiently deep.

In order to investigate the flux limits at which confusion noise dominates 
over Poisson errors, we have produced and source-searched a small number of 
ultra-deep simulations with no Poisson noise. The results are presented and 
discussed in 
Appendix B, and show that the \hr~ flux limits are more than a factor 
of 4 brighter than the ultimate \xmm confusion limit in any of the 4 energy 
bands. 

Eddington bias \citep{eddington1913}
results in a systematic offset in the number of sources detected
at a given flux. The magnitude of this effect depends on the both the
statistical errors on the measured flux values and on the intrinsic slope of
the \ns$\!$. As there are generally many more faint sources than bright ones,
uncertainties on the measured flux values will result in more faint
sources being up-scattered  than bright sources being
down-scattered. Therefore we would expect more faint sources to be detected
than are actually input, and the output source counts at a given flux to be
greater than those input. In the case
of our simulations the situation is further complicated due to the double
powerlaw form of the \ns distribution and the fact that the
flux error distribution is non-uniform and a function of several parameters
including flux and offaxis angle.

\begin{figure}
\setlength{\unitlength}{1in}
\begin{picture}(3.5,3.5)
\put(-0.1,0.2){\includegraphics{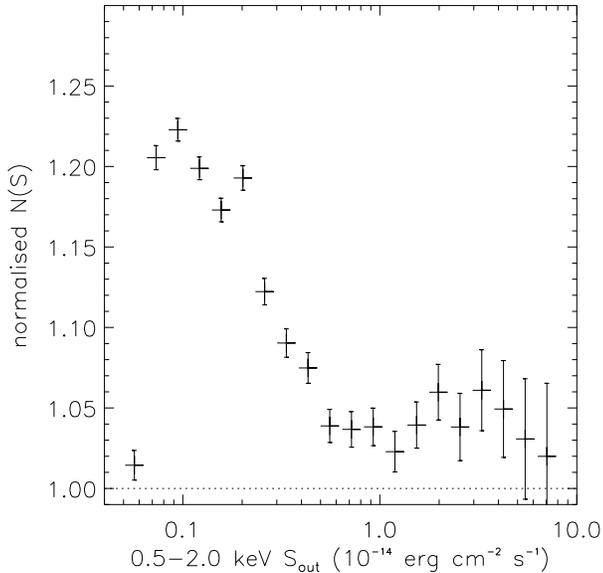}}
\end{picture}
\vspace{-0.8cm}
\caption{Simulated output (crosses) 0.5-2 keV \ns distribution normalised 
by the input distribution. The error bars represent Poisson errors on the
output source counts. The dotted line represents the case where the input and
output source counts are equal. The source counts are most disparate at faint
fluxes where the output counts are enhanced by up to 23\%.}
\label{fig:lowsim}
\end{figure}

The level of Eddington bias expected in the \hr~ deep field is shown in
Fig. \ref{fig:lowsim} where the simulated input and output source counts are
compared. Below $S_{knee}$ the ratio rises
as the statistical errors on the flux measurements increase. 
At the lowest flux interval there is a drop in output source counts. 
The reason for this is the strong skew in $S_{out}/S_{inp}$ at
the faintest fluxes (see Fig. \ref{fig:hist}) which boosts the output fluxes. 
The output/input source counts ratio is a minimum at $S_{knee}$ where both the 
statistical errors on the flux measurement are low and the source counts become flat. 
Above $S_{knee}$ the output/input source count ratio is constant 
within the errors with a value of $\sim1.05$ indicating that Eddington bias
affects our bright counts at the 5\% level.

\subsection{Assessing the reliability of the \hr~  sources}
\label{sec:likes}
In this section we describe how we determined from our simulations 
the appropriate detection threshold
to be applied to the \hr~ field EMLDETECT sourcelist. Our aim was to produce a 
sourcelist such that nearly all erroneous detections are removed whilst retaining the 
maximum number of real sources.

For each detected source, EMLDETECT measures the detection maximum likelihood
statistic, DET\_ML, which takes account of source counts, background counts,
and the PSF. 
For the real data, the SAS task EMLDETECT provides both single-band and
multi-band measurements of DET\_ML. For the simulated images we are limited to
single band measurements as the four bands are simulated independently.
To simulate all four bands simultaneously would require \emph{a priori}
knowledge of the sources' intrinsic X-ray spectra, redshifts and column
densities which we do not have.
The value of DET\_ML for any single detection is directly related to the
probability of the source being caused by a random Poisson fluctuation 
via Eqn. \ref{eqn:cash}.
However, it is difficult to translate a minimum threshold value of 
DET\_ML applied to
the whole sourcelist into a total number of expected spurious sources in the
field as the probability is a function of position within the field, due to the
varying PSF and exposure map. This is therefore best explored via a large
number of Monte-Carlo simulations.

\begin{figure}
\setlength{\unitlength}{1in}
\begin{picture}(3,3.5)
\put(-0.6,-2.5){\includegraphics{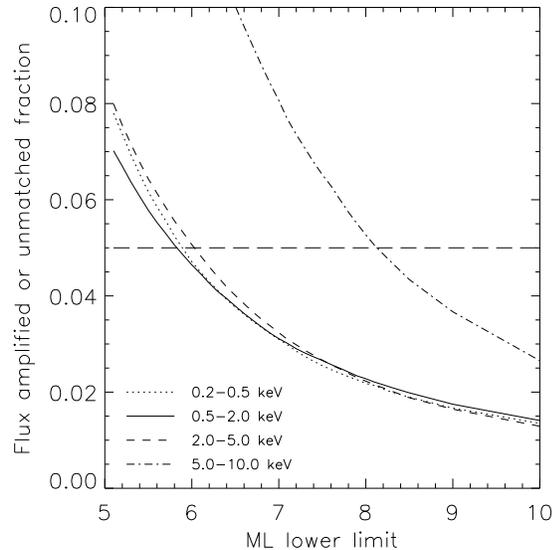}}
\end{picture}
\vspace{-0.5cm}
\caption{Plot showing the fraction of flux amplified or unmatched sources for a range of
lower DET\_ML limits. Results are shown for each energy band, 0.2-0.5keV
(large dash), 0.5-2keV (solid line), 2-5keV (small dash), 5-10keV
(dot-dash). The dotted line shows the 5\% badness level. }
\label{fig:single_band_badness}
\end{figure}

Using the simulated sourcelists in each band, we calculated the 
fraction of sources which were either flux amplified or unmatched as a function
of the minimum detection threshold $\rm DET\_ML_{min}$. This is shown in 
Fig. \ref{fig:single_band_badness}. 
To restrict the fraction of bad sources in our final \hr~ sourcelist 
we use the value of $\rm DET\_ML_{min}$ 
in each band at which only 5 percent of sources are either flux amplified or
unmatched in our simulations; these are 5.9,
5.9, 6.0, 8.1 in the 0.2-0.5, 0.5-2, 2-5, and 5-10 keV bands respectively.
For a source to pass our significance threshold we require that it has
$\rm{DET\_ML \ge DET\_ML_{min}}$ in \emph{at least} one energy band. 
In practice, because we
use a multi-band source detection process for the \hr~ data, we expect  
fewer than 5\% bad sources after we have applied this criterion, 
as many of the sources will be detected in more than one energy band. 

For each source in
the \hr~ sourcelist, we can identify all the output sources from the simulations
which lie within 2\arcmin~ of the real \hr~ source and have a similar DET\_ML (within 10\%). 
The fraction of these output sources which are unmatched or flux amplified
gives a good estimate of the probability that the detection of the real source 
in this energy band is unreliable. 
For sources detected in more than one band 
the probability that the source is spurious is given by the product of these 
individual probabilities. The total number of spurious sources expected in the
\hr~ field can then be estimated by summing the probabilities from each source. 

\begin{figure}
\setlength{\unitlength}{1in}
\begin{picture}(3,3.4)
\put(-0.5,-2.5){\includegraphics{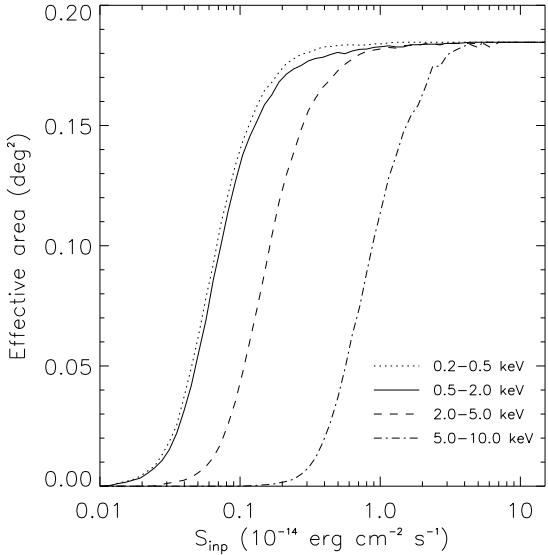}}
\end{picture}
\vspace{-0.5cm}
\caption{Effective area for each energy band, determined from Monte-Carlo
simulations (0.2-0.5 keV (dotted), 0.5-2 keV (solid), 2-5keV (dashed), 5-10
keV (dot-dashed)).}
\label{fig:effarea}
\end{figure}


\subsection{Maximum likelihood \ns fitting method}
\label{sec:murdoch}

Our simulations are ideal for testing our \ns fitting procedure because
we know \emph{a priori} the input \ns fitting parameters.
Accurate fitting of any \ns relation requires a knowledge of the
sky area searched and the probability of detection at a given flux.
In  Fig. \ref{fig:effarea} the effective area of the survey as a function of 
limiting flux is shown for each energy band. This was determined by comparing 
the number of output to input sources at each flux and multiplying the resultant
fraction by the geometric area of the survey. 
Additionally, the distribution of $S_{out}/S_{inp}$  is required to account 
for Eddington bias (see Fig. \ref{fig:hist}, and Section \ref{sec:simcomp}). 

Adapting the maximum likelihood method of 
\citet{murdoch73}, we fitted a double powerlaw model to the output \ns
relation. Rather than fitting  the observed
fluxes directly, we convolved a model \ns with the distribution of
$S_{out}/S_{inp}$  
to produce a model probability distribution of observed fluxes
$P(S_{out})$. The best-fit \ns was then determined by maximising the likelihood of 
obtaining the observed flux distribution.
As a check, we applied this technique to our simulated data, and recover the
input values to within the statistical limits of our simulations ($\sim$2\%).
We are therefore confident in applying this technique to the real \hr~ data.
Fits neglecting the distribution of $S_{out}/S_{inp}$  are typically 
systematically offset by $\sim$5\%

\section{Results}
\label{results}

In this section we present our final source catalogue and results of fits to
our source counts.
We compare the properties of our sources with those detected by \chandra in the
same area and closely examine our quoted positional uncertainties via comparisons 
with both \chandra detections and our simulations.

\subsection{X-ray source catalogue}

Following the source detection procedure described in Section
\ref{sec:background} (simultaneous source searching in four energy bands), 
a total of 275 sources were detected with $ \rm {DET\_ML > 5}$ in
at least one energy band. This sourcelist
was curtailed using the $\rm {DET\_ML_{min}}$ values for each energy band
determined in Section \ref{sec:likes}.
This reduced the final number of sources to 225. 
The final sourcelist is presented in Table \ref{tab:point_sources}. 
Using the procedure described in Section \ref{sec:likes} we expect 
a total of 7 spurious sources.

\subsection{\hr~ deep field source counts}
\label{sec:srccounts}
\begin{table}
\begin{center}
\begin{tabular}{lcrl}
\hline \hline
Energy (keV) & \multicolumn{1}{c}{${\gamma}$} & \multicolumn{1}{c}{$K$} & KS Prob\\
\hline
0.2--0.5 & $1.84^{+0.20}_{-0.18}$ &   47$\pm$20 & 1.8$\times10^{-2}$\\
0.5--2.0  & $1.69^{+0.11}_{-0.11}$ &  112$\pm$23 & 1.0$\times10^{-10}$\\
2.0--5.0  & $1.91^{+0.20}_{-0.19}$ &  126$\pm$28 & 3.5$\times10^{-4}$\\
5.0--10.0 & $2.80^{+0.67}_{-0.55}$ &  150$\pm$18 & 2.9$\times10^{-5}$\\
\hline
\end{tabular}
\caption{ Best-fit values for a single powerlaw fit to the \hr~ deep field
differential source counts. All errors are at 95\% confidence. The best-fit
slope, $\gamma$, is listed together with the normalisation, K, in units of
$ \rm {(10^{-14}~erg~cm^{-2}~s^{-1})^{\gamma-1}~deg^{-2} }$.  KS null-hypothesis
probabilities on the fits are listed in the final column. }
\end{center}
\label{tab:singleresults}
\end{table}

\begin{figure*}
\setlength{\unitlength}{1in}
\begin{picture}(6.0,7.0)
\put(-0.5,0.5){\includegraphics{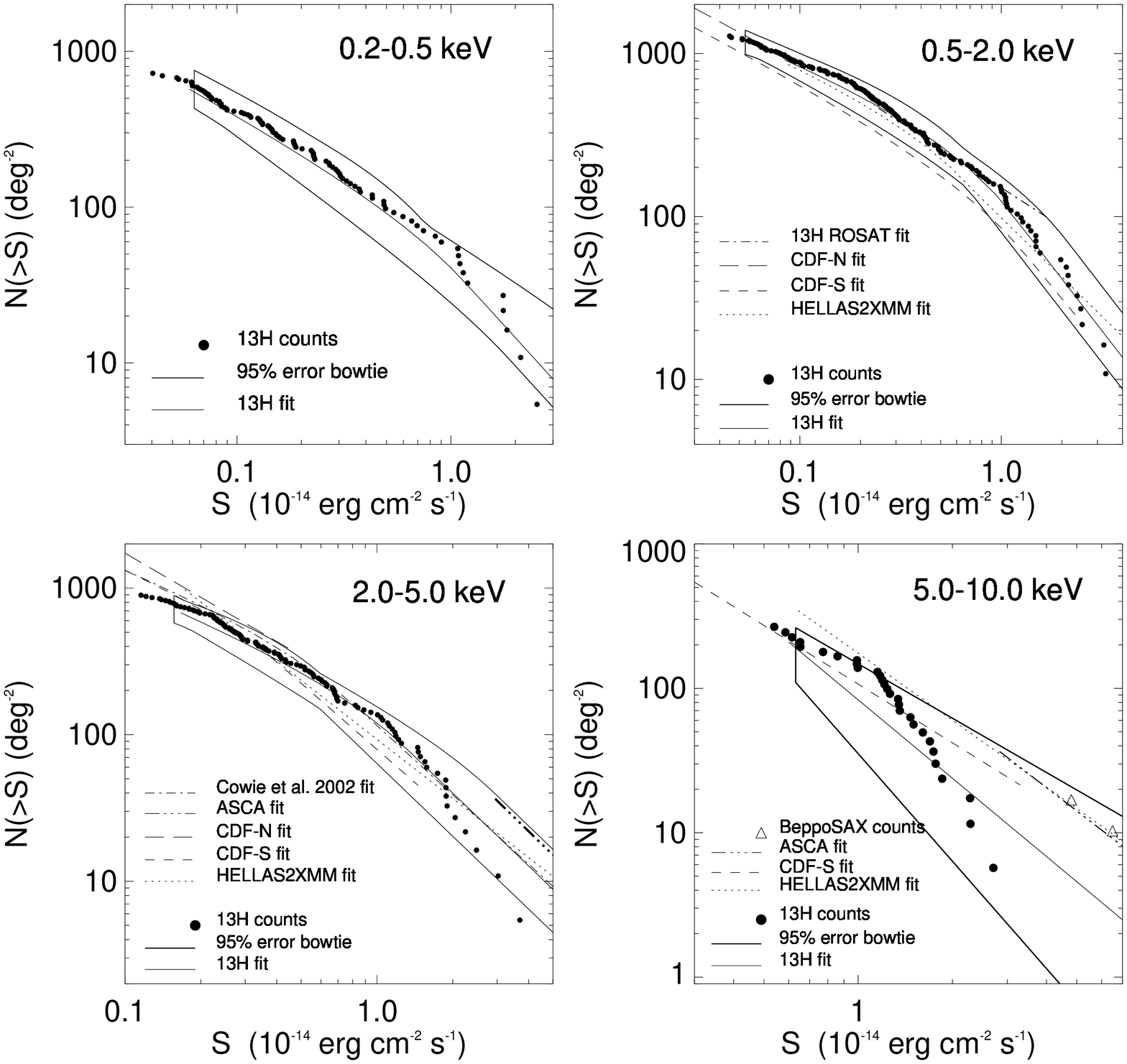}}

\end{picture}
\vspace{-1.0cm}
\caption{\hr~ deep field integral \logns in each energy band. Bowties indicate
95\% errors. Overlaid for comparison are the results from the CDF-S (dash)
\citep{rosati02}, CDF-N (long dash) \citep{brandt01}, HELLAS2XMM (dot)
\citep{baldi02} and \asca (dot-dot-dot-dash) \citep{cagnoni98} surveys. 
In the 0.5--2.0 keV band the ROSAT counts found in the \hr~ field
\citep{mchardy98} are also overlaid (dot-dash).
In the
2--5 keV band the results from \citet{cowie02} are also overlaid (dot-dash).
Triangles in the 5--10 keV band  denote the \sax counts of \citet{fiore01}.}
\label{fig:lognlogs}
\end{figure*}

We show  the \emph{integral} \logns in each
energy band in Fig. \ref{fig:lognlogs}. 
We have fitted single and double powerlaw models to the unbinned
differential source counts in all energy bands using the method described in
Section \ref{sec:murdoch}. 
The four identified Galactic stars in the field have been excluded from 
this analysis.   
The best-fit double powerlaw models are overlaid in Fig. \ref{fig:lognlogs}
(single powerlaw in the 5-10 keV energy band) as solid lines. 
The 95\% confidence interval of the fits are indicated by the bowties. These
incorporate  errors on the slopes, normalisations and knee if applicable.

The best-fit
parameters for a single powerlaw fit are listed in Table
\ref{tab:singleresults}. 
The uncertainties on the fit parameters are quoted at the 95\% confidence 
interval for one interesting parameter.
We have tested the goodness of fit of these models
using the Kolmogorov-Smirnov (KS) test, and the null-hypothesis probabilities
that we obtain are given in Table \ref{tab:singleresults}. 
The fits are unacceptable in all bands.

For the double powerlaw fits we have fixed the slope at bright fluxes
($\gamma_{2}$) and fit only for the slope at faint fluxes ($\gamma_{1}$) and
knee position ($S_{knee}$).
In the 0.2-0.5 keV band we set $\gamma_{2}=2.51$ based on results from the 
\rosat All Sky Survey \citep{voges99}. The value of $\gamma_{2}$ in the
0.5-2 keV band was fixed at 2.60, derived from the RIXOS survey
\citep{mason2000}. 
For the 2-5 keV band $\gamma_{2}$ was fixed at 2.65 as found in the \sax High Energy
Large Area Survey \citep[HELLAS][]{giommi00}, and consistent with
the \asca derived value of 2.5 from \citet{ueda99}. 
The fit parameters for the double powerlaw fits are listed in Table 
\ref{tab:doublefits}. 
The double powerlaw model provides a better fit compared to a single powerlaw model in
each of these three energy bands.  
However, the double powerlaw model
fit is formally rejected with 99\% confidence in the 0.5-2 and 2-5 keV energy bands.
In all three energy bands $S_{knee}$ occurs between $1.08-1.27\times10^{-14}$ \ergs$\!\!$. 

In the 5-10 keV band the
slope found in the single powerlaw fit is consistent with that found by
\citet{baldi02} at brighter fluxes. We attempted to fit a double powerlaw
model to the 5-10 keV band source counts, but in this case $S_{knee}$ is
unconstrained. We therefore consider that a double powerlaw model for the
source counts is not justified at the depth of our survey in this energy band.

\begin{table*}
\begin{center}
\begin{tabular}{lcllrrr}
\hline \hline
Energy (keV) & ${\gamma _1}$ & ${\gamma _2}$ &
\multicolumn{1}{c}{${S_{knee}}$} & \multicolumn{1}{c}{$K_{1}$} &
\multicolumn{1}{c}{$K_{2}$} & KS Prob \\
\hline
0.2--0.5  & $1.74^{+0.25}_{-0.26}$ &  2.51 &  $1.16^{+11.08}_{-0.69}$ &   56$\pm$23
&   62$\pm$88 & 7.7$\times10^{-2}$\\
0.5--2.0 & $1.41^{+0.19}_{-0.18}$ &  2.60  &  $1.08^{+1.02}_{-0.39}$ &  183$\pm$51
&  201$\pm$67 & 6.4$\times10^{-5}$\\
2.0--5.0 & $1.66^{+0.30}_{-0.47}$ &  2.65 &  $1.27^{+1.66}_{-0.70}$ &  163$\pm$94 &
 207$\pm$76 & 1.5$\times10^{-2}$\\

\hline
\end{tabular}
\caption{Best-fit parameters for a double powerlaw fit to the \hr~ differential
source counts. The slopes and normalisations below and above the break flux,
$S_{knee}$ are denoted by $_1$ and $_2$ respectively.
The values of $\gamma_{2}$ in the three energy bands were
fixed to appropriate values found from the literature. The break flux,
${S_{knee}}$ is in units of $10^{-14}$ \ergs$\!\!$. 
Normalisations $K_{1}$ and $K_{2}$ are in units of 
$(10^{-14}~erg~cm^{-2}~s^{-1})^{\gamma-1}~deg^{-2}$. 
All errors are at 95\%. 
KS null-hypothesis probabilities on the fits are listed in the final column.}
\label{tab:doublefits}
\end{center}
\end{table*}

\subsection{Positional uncertainties}
\label{sec:comparison}

To assess the reliability of the EMLDETECT positional
errors in the \xmm sourcelist we have cross-correlated the final \xmm and
\chandra sourcelists. The spread in positional offsets between the two
catalogues should provide a good representation of the true spread in \xmm
positional errors. 
To prevent the (albeit small) statistical errors on the \chandra positions contributing to
the spread, where possible we used the position of the optical
counterpart to the \chandra source: the positional errors on the optical counterparts 
are typically less than 0.3\arcsec~ \citep{mchardy03}. 
The two sourcelists were matched within a search radius of
10\arcsec~ over the entire \xmm field of view.  

\begin{figure}
\setlength{\unitlength}{1in}
\begin{picture}(3.5,3.5)
\put(-0.5,-2.5){\includegraphics{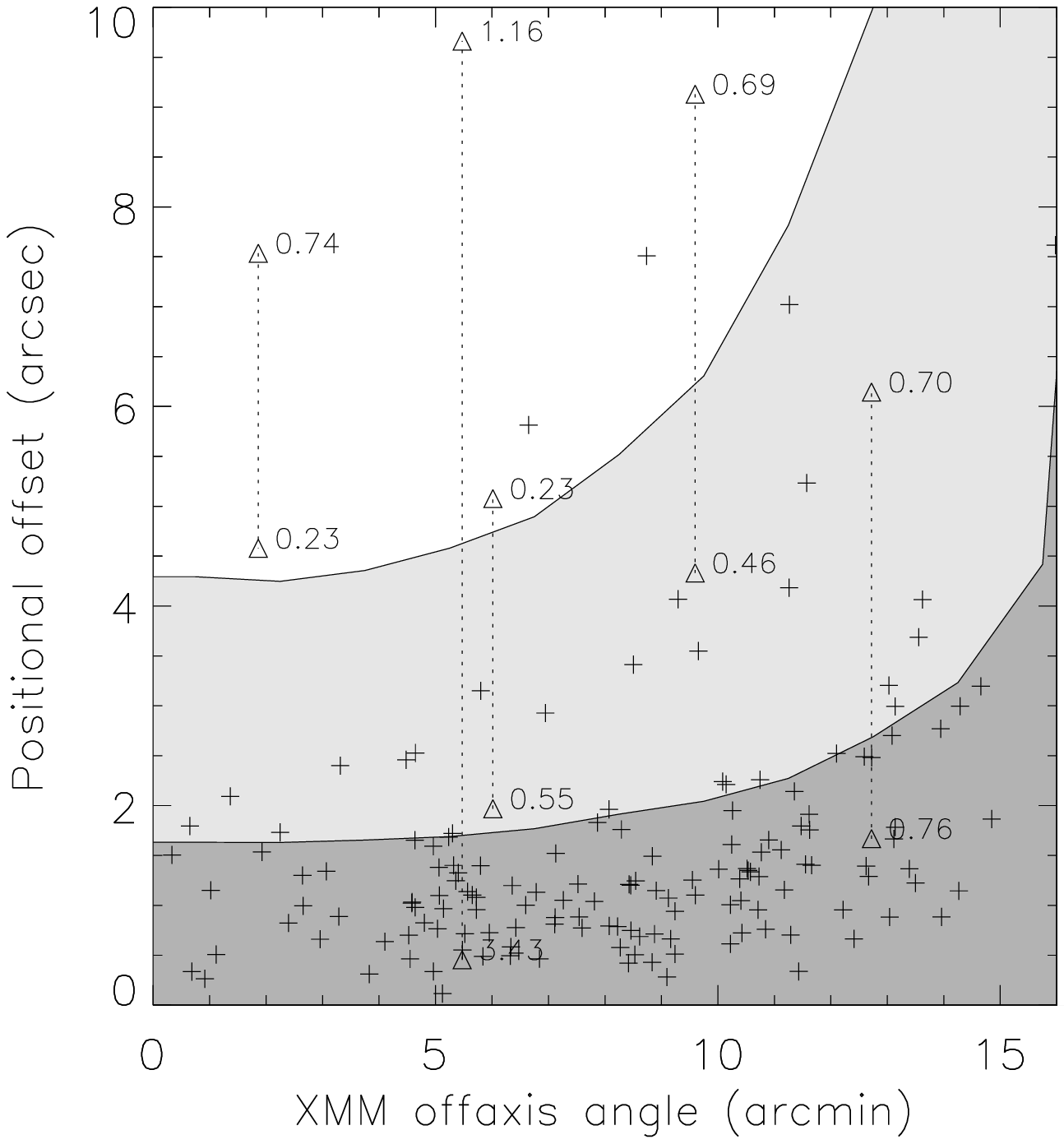}}
\end{picture}
\vspace{-1.0cm}
\caption{Positional offsets between \xmm and \chandra counterparts as a
function of \xmm offaxis angle. Triangles denote \xmm detections with $>1$
\chandra counterparts within 10\arcsec~ (numbers indicate the \xmm flux in
units of $10^{-14}$ \ergs). The shaded areas indicate the regions
where 68\% (dark grey) and 95\% (light grey) of the simulated data lie. In the
case of the simulations the positional offset is that between the input and
output source. 86\% of the \hr~ data lies within the 68\% simulation
contour, indicating that the real data have better source positions. This is
due to the fact that the real data are generally detected in $>1$ energy band. }
\label{fig:comparisons}
\end{figure}

The two catalogues have 155 sources in common.
The positional offset between the 155 matched \xmm and \chandra sources as a
function of \xmm offaxis angle is shown in Fig. \ref{fig:comparisons}. 
The positional offsets increase with \xmm offaxis angle, but
80 per cent of the matched sources have positional offsets $\leq 2\arcsec$; 
95 per cent of the sources have positional offsets $ \leq 4\arcsec$. 

We compared this distribution with that found from the 
simulations (see Section \ref{sec:simcomp}) which should provide a good
indication of the distribution of positional offsets arising within the
source detection chain.
The contours in Fig. \ref{fig:comparisons} show the distribution of 
differences between input and output source positions in the 0.5-2 keV
simulations. 
The dark (light) grey area shows where 68\% (95\%) of the simulated sources
lie. The distribution of \xmm-\chandra offsets is more highly peaked than the
input-output offsets in
our simulations: 86\% of \xmm-\chandra offsets lie within the 68\% simulation
contour.
This is because the real sources in the \hr~ field are generally detected in
more than one energy band, contrary to the simulated sources. Fitting the source positions
in all four energy bands simultaneously, as was done to produce our source
catalogue, results in better positional accuracy than a single band fit because
more source counts are used in the fit.

\begin{figure}
\setlength{\unitlength}{1in}
\begin{picture}(3.5,3.5)
\put(-0.5,-2.5){\includegraphics{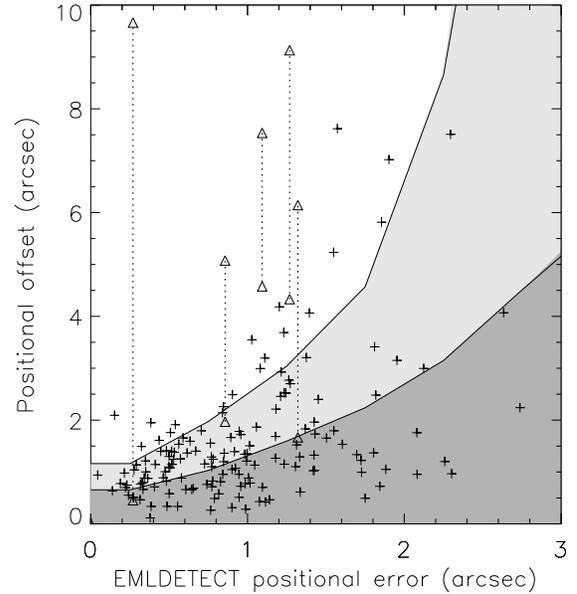}}
\end{picture}
\vspace{-1.0cm}
\caption{ Positional offsets between \xmm and
\chandra counterparts compared with quoted \xmm positional error. The shaded 
areas indicate the regions
where 68\% (dark grey) and 95\% (light grey) of the simulated data lie. In the
case of the simulations the positional offset is that between the input and
output source. 43\% of the \hr data lie within the 68\% simulation contour
indicating that EMLDETECT underestimates the true positional error.}
\label{fig:comparisons2}
\end{figure}

Five of the matched sources have more than one \chandra counterpart. However,
in two out of the five cases, the closest counterpart is significantly brighter
than the other one, and we are easily able to identify the correct
counterpart.  This suggests that $\sim2$\% of our sources are confused, in
broad agreement with the fraction of flux amplified sources expected from our
simulations.

In Fig. \ref{fig:comparisons2} we plot the statistical positional
uncertainties of the \xmm sources given by EMLDETECT against the actual
positional uncertainties given by the \xmm - \chandra offsets. These are
compared with our 0.5-2 keV simulation distributions where in the case of the
simulations the positional offsets are those of the input/output source
positions. The dark (light) grey areas show where 68\% (95\%) of the simulated
sources lie. In general, the positional offsets are slightly larger than the
EMLDETECT errors for the \hr~ data.
However, for EMLDETECT positional errors less than 1\arcsec~ the positional
offsets have a broader distribution in the \hr~ data than in the simulations: 
only 43\% of the \hr~ data lies within the 68\% simulation region. 
This implies that the EMLDETECT positional
error in the \hr~ field underestimates the true positional error. We have
therefore added in quadrature a systematic error to the statistical EMLDETECT positional
error of each \hr~ source, such that 68\% of the
\hr~ sources lie within the 68\% region of the simulations.
This is achieved with an additional systematic positional error of
0.76\arcsec~($1\sigma$). This additional component may be due to residual 
uncertainties in the detector geometry and may represent a fundamental limit 
to the  accuracy of \emph{any} \xmm position.  
According to our simulations, EMLDETECT positional errors
greater than $\sim2.5$\arcsec~ may underestimate the true positional error by
an order of magnitude (the 95\% simulation region limit samples a non-gaussian
tail at this point) and should be used with this caveat in mind.
Section \ref{sec:reliab} further discusses the reliability of the survey and the
implications of the positional uncertainties.

\begin{figure*}
\setlength{\unitlength}{1in}
\begin{picture}(5.0,8.5)
\put(-0.7,2.5){\includegraphics{rpics1.ps}}
\put(-0.7,0.25){\includegraphics{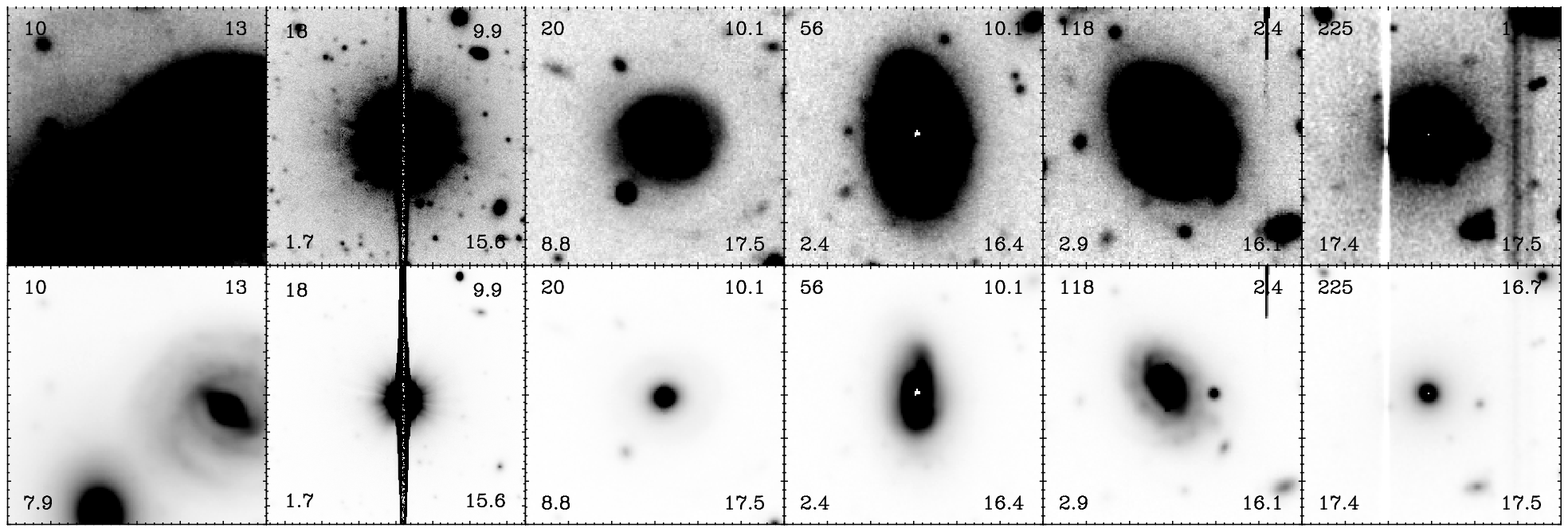}}
\end{picture}
\vspace{-0.7cm}
\caption{Top: $10" \times 10"$ $R$-band Subaru SuprimeCam images centred on the
\xmm sources without \chandra counterparts.  The \xmm  reference number 
(shown in column 1 of Table \ref{tab:point_sources}) is at the 
top left of each image. The \xmm off-axis angle is at the top right. The 
\xmm $0.2-10$ keV flux (in units of $10^{-15}$ \ergs) is at the bottom left and
the optical counterpart R magnitude, if applicable, is given in the bottom
right.  All of these SuprimeCam images have the same greyscale levels to ease 
comparison of the optical counterparts' brightnesses. The greyscale was chosen
to correspond to that in \citet{mchardy03}.
Bottom:
$30\arcsec \times 30\arcsec$~ $R$-band Subaru SuprimeCam images centred on the 
6 \xmm sources with \chandra counterparts which are too extended to fit in a
$10\arcsec \times 10\arcsec~$ box. Source number 18 is actually shown in a
$70\arcsec \times 70\arcsec~$ box. The top images are at the same grey scale
level as Fig. \ref{fig:optimage}. The lower images have their grey scales
adjusted for maximum contrast. Source number 10 is not listed as having an optical
counterpart in Table \ref{tab:xmmnochandra} as the
centre of the barred spiral galaxy is outside the error circle of the \xmm
source. However, we may be seeing X-ray emission from an X-ray source
in the spiral arms of the galaxy.}
\label{fig:optimage}
\end{figure*}

\begin{table*}
\begin{center}
\scriptsize
\begin{tabular}{rccrrrrcccl}
\hline \hline
\multicolumn{1}{c}{ \xmm } &
\multicolumn{1}{c}{RA J(2000) } & \multicolumn{1}{c}{Dec J(2000)} &
\multicolumn{1}{c}{Pos. err.} & \multicolumn{1}{c}{Offaxis} &
\multicolumn{1}{c}{0.2-10.0 keV} & 
\multicolumn{1}{c}{Extension} &
\multicolumn{1}{c}{R mag.} &
\multicolumn{1}{c}{\chandra} & \multicolumn{1}{c}{Faint \chandra} 
& \multicolumn{1}{c}{Comments} \\
\multicolumn{1}{c}{ No. } & & & 
\multicolumn{1}{c}{(arcsec)} & \multicolumn{1}{c}{(arcmin)} &
\multicolumn{1}{c}{Flux} & & &
\multicolumn{1}{c}{counts} & \multicolumn{1}{c}{source?} \\
\hline
1  &  13 33 28.96 & 37 55 58.82 &  1.57 &  12.91 &   14.86  &   -- & 19.74 &  3 &  n  \\
2  &  13 33 30.32 & 37 56 12.21 &  2.38 &  12.67 &    5.94  &   -- & 23.18 &  1 &  n  \\
3  &  13 33 30.68 & 37 55 05.57 &  3.04 &  12.53 &    2.55  &   -- & 25.83 &  0 &  n  \\   
10 &  13 33 38.91 & 38 01 56.15 &  4.83 &  12.96 &    7.88  &   -- &  0.00 &  0 &  n  \\
11 &  13 33 40.70 & 37 49 47.79 &  1.99 &  11.74 &    1.65  &   -- & 23.76 &  2 &  n  \\
18 &  13 33 43.80 & 37 54 55.76 &  2.56 &   9.95 &    1.74  &   -- & 15.63 &  3 &  y  \\
20 &  13 33 45.45 & 37 58 08.04 &  0.92 &  10.14 &    8.82  &   -- & 17.45 &  0 &  n  \\  
22 &  13 33 47.52 & 37 53 51.51 &  1.11 &   9.27 &    3.46  &   -- & 23.39 &  1 &  n  \\  
26 &  13 33 50.80 & 37 57 16.61 &  1.77 &   8.88 &    4.43  &   -- & 25.41 &  3 &  y \\
28 &  13 33 54.21 & 38 02 53.17 &  1.23 &  11.21 &   16.49  &   -- & 25.41 &  0 &  n \\ 
31 &  13 33 57.90 & 37 49 59.01 &  2.91 &   8.70 &    0.98  &   -- & 24.79 &  1 &  n \\ 
33 &  13 33 58.87 & 38 00 24.88 &  1.18 &   8.87 &    6.60  &   -- & 24.83 &  1 &  n \\ 
35 &  13 33 59.56 & 37 51 39.75 &  2.34 &   7.57 &    2.06  &   -- & 25.69 &  2 &  n \\ 
36 &  13 33 59.63 & 37 49 26.35 & 12.19 &   8.75 &    1.86  &   -- & 17.64 &  0 &  n \\ 
40 &  13 34 01.06 & 38 00 25.75 &  1.50 &   8.55 &    5.92  &   -- & 20.50 &  1 &  n \\ 
44 &  13 34 07.18 & 37 59 59.34 &  1.23 &   7.36 &    4.87  &   -- & 22.50 &  0 &  n \\  
47 &  13 34 08.39 & 37 47 50.85 &  2.69 &   8.72 &    4.67  &   -- & 22.65 &  1 &  n \\ 
54 &  13 34 12.36 & 37 59 10.36 &  2.01 &   6.06 &    1.10  &   -- & 26.57 &  1 &  n \\ 
56 &  13 34 13.59 & 37 45 39.06 &  1.22 &  10.12 &    2.42  &   -- & 16.43 &  3 &  y \\
62 &  13 34 15.49 & 38 03 05.78 &  1.80 &   8.97 &    4.81  &   -- & 19.68 &  2 &  n \\
63 &  13 34 15.54 & 37 52 27.80 &  1.31 &   4.43 &    5.54  &   -- & 23.65 &  2 &  y \\
66 &  13 34 17.80 & 37 44 31.39 &  2.04 &  10.89 &    3.51  &   -- & 18.76 &  2 &  y \\ 
70 &  13 34 19.56 & 37 51 47.77 &  1.40 &   4.26 &    3.58  &   -- & 22.89 &  0
&  y  &  Possible faint source outside error circle\\
79 &  13 34 22.70 & 37 55 23.49 &  2.15 &   2.32 &    0.90  &   -- & 18.17 &  3 &  n \\ 
82 &  13 34 25.91 & 37 54 59.90 &  1.58 &   1.64 &    1.63  &   -- & 23.94 &  2 &  n \\ 
83 &  13 34 26.87 & 38 00 27.92 &  1.45 &   5.73 &    2.66  &   -- &  0.00 &  1 &  n \\ 
87 &  13 34 28.66 & 37 57 48.26 &  1.92 &   3.09 &    2.21  &   -- & 20.64 &  3 &  y \\
88 &  13 34 28.80 & 37 53 37.96 &  1.78 &   1.67 &    2.15  &   -- & 18.46 &  0 &  n \\  
89 &  13 34 29.20 & 38 02 45.33 &  1.79 &   7.90 &    0.89  &   -- & 22.56 &  0 &  n \\ 
90 &  13 34 29.30 & 38 06 51.45 &  1.76 &  11.98 &    7.29  &   -- & 21.88 &  1 &  n &   Edge of \chandra FOV\\
104&  13 34 34.92 & 38 07 03.03 &  1.99 &  12.13 &    6.71  &   -- & 24.47 &  3 &  n \\ 
110&  13 34 36.38 & 38 05 14.02 &  1.25 &  10.32 &    5.16  &   -- & 25.32 &  4 &  y \\
116&  13 34 38.83 & 37 40 22.03 &  1.11 &  14.58 &   34.78  &   -- & 25.30 &  0 &  n &   Outside \chandra FOV  \\ 
117&  13 34 38.96 & 37 42 43.95 &  1.60 &  12.22 &    5.49  &   -- & 24.30 &  0 &  n \\   
118&  13 34 39.75 & 37 57 01.93 &  1.29 &   2.38 &    2.91  &   -- & 16.14 &  4 &  y \\
120&  13 34 42.29 & 37 41 46.46 &  0.78 &  13.24 &   10.84  &   -- & 20.14 &  0 &  n &   Outside \chandra FOV \\
122&  13 34 42.81 & 37 42 42.62 &  1.55 &  12.33 &    9.25  &   -- &  0.00 &  0 &  n &   Edge  of \chandra FOV \\
123&  13 34 42.84 & 38 03 53.27 &  2.33 &   9.13 &    1.65  &   -- & 18.53 &  1 &  n \\ 
133&  13 34 47.18 & 37 57 14.88 &  1.53 &   3.46 &    3.71  &   -- & 23.73 &  4 &  y \\
136&  13 34 48.22 & 37 54 14.33 &  1.06 &   2.85 &    7.93  &   -- &  0.00 &  0 &  n \\  
139&  13 34 51.27 & 37 40 51.30 &  1.23 &  14.46 &   17.14  &   -- &  0.00 &  0 &  n &   Outside \chandra FOV\\
141&  13 34 51.86 & 38 06 35.33 &  1.78 &  12.18 &    4.16  &   -- & 22.80 &  1 &  n \\    
144&  13 34 53.13 & 37 48 18.85 &  1.76 &   7.59 &    3.59  &   -- & 24.06 &  2 &  y  \\
147&  13 34 54.55 & 37 42 07.32 &  1.30 &  13.41 &   12.44  &   -- & 0.00  &  4 &  n \\ 
149&  13 34 55.11 & 37 49 51.41 &  2.44 &   6.53 &    2.87  &   -- & 25.26 &  1 &  n \\
150&  13 34 55.41 & 37 50 40.54 &  2.66 &   5.96 &    1.32  &   -- & 24.95 &  4 &  n \\ 
151&  13 34 56.39 & 37 39 54.75 &  2.06 &  15.63 &   15.15  &   -- & 24.89 &  0 &  n &   Outside \chandra FOV  \\ 
153&  13 34 56.71 & 37 50 25.09 &  1.74 &   6.32 &    4.97  &   -- & 0.00  &  0 &  n \\ 
154&  13 34 56.76 & 37 56 03.31 &  1.48 &   4.59 &    3.04  &   -- & 20.66 &  0 &  n \\ 
155&  13 34 56.87 & 37 52 48.73 &  2.34 &   4.94 &    0.63  &   -- & 25.00 &  0 &  n \\ 
159&  13 34 58.11 & 37 57 33.15 &  1.22 &   5.40 &    3.34  &   -- &  0.00 &  0 &  n \\ 
160&  13 34 58.14 & 37 50 52.22 &  2.32 &   6.22 &    1.16  &   -- &  0.00 &  1 &  n \\ 
163&  13 34 59.48 & 37 57 39.65 &  1.17 &   5.69 &    7.07  &   -- & 26.38 &  2 &  n \\ 
172&  13 35 05.13 & 37 45 27.22 &  1.90 &  11.26 &    3.52  &   -- &  0.00 &  0 &  n \\ 
177&  13 35 07.69 & 37 48 27.88 &  2.66 &   9.24 &    2.74  &   -- & 25.23 &  0 &  n \\  
179&  13 35 09.25 & 38 04 01.91 &  1.55 &  11.43 &    6.08  &   -- &  0.00 &  1 &  n \\ 
182&  13 35 12.27 & 37 48 54.89 &  1.86 &   9.62 &    1.73  &   -- & 24.32 &  0 &  n \\   
185&  13 35 14.41 & 37 49 12.11 &  1.49 &   9.78 &   20.46  &   9.2$\pm$0.4  & 20.55 &  0 &  n \\   
194&  13 35 17.03 & 37 49 13.98 & 14.06 &  10.19 &    1.44  &   -- & 19.44 & 15 &  n &  Large error circle \\  
205&  13 35 24.72 & 37 51 24.06 &  1.68 &  10.57 &    3.14  &   -- &  0.00 &  0 &  n \\ 
206&  13 35 24.88 & 37 44 38.51 &  1.90 &  14.34 &    3.94  &   -- & 23.41 &  0 &  n \\ 
209&  13 35 32.22 & 38 03 39.97 &  0.70 &  14.39 &  238.21  &   62$\pm$104 & 25.10 &  0 &  n \\  
210&  13 35 32.74 & 37 45 14.16 &  4.66 &  15.08 &   71.49  &   26$\pm$2 & 22.43 &  2 &  n \\ 
211&  13 35 33.05 & 37 48 02.14 &  1.98 &  13.50 &    3.24  &   -- & 18.54 &  1 &  n \\ 
215&  13 35 37.26 & 37 47 22.80 &  1.34 &  14.55 &    5.96  &   -- & 20.28 &  0 &  n \\  
217&  13 35 39.37 & 37 56 14.96 &  0.68 &  12.92 &   20.15  &   -- & 20.89 &  0 &  n &   Outside \chandra FOV \\  
218&  13 35 42.49 & 37 55 42.93 &  0.59 &  13.49 &   23.77  &   -- & 19.47 &  0 &  n &   Outside \chandra FOV \\  
219&  13 35 43.17 & 37 52 58.65 &  1.72 &  13.74 &    6.49  &   -- & 26.58 &  1 &  n &   \chandra 102 is $\sim10"$ North \\
220&  13 35 43.26 & 37 53 24.27 &  2.17 &  13.70 &    3.95  &   -- & 25.89 &  0 &  n &   \chandra 102 is $\sim15"$ South \\
225&  13 35 58.70 & 37 54 06.31 &  2.76 &  16.68 &   17.44  &   6.8$\pm$1.2 & 17.51 &  0 &  n &   Outside \chandra FOV   \\
\hline 
\end{tabular}
\caption{Properties of \xmm sources not found in the \chandra catalogue of
\citet{mchardy03}. The \xmm flux is in units of $10^{-15}$ \ergs$\!\!$. The extension
of the source (if applicable) in the \xmm image in units of arcsec.
The column marked `\chandra counts' gives the  number of counts detected within the
\xmm error circle in the \chandra image. Excess emission associated with a
possible faint source is indicated by a `y' in column 10. See the main text for
further details.}
\label{tab:xmmnochandra}
\end{center}
\end{table*}

\subsection{Comparisons with \chandra detections}
\label{sec:nochandra}

\begin{table*}
\begin{center}
\begin{tabular}{ccccl}
\hline \hline 
\chandra & \xmm offaxis angle & 0.5-7 keV  & \multicolumn{1}{c}{Faint \xmm} & \multicolumn{1}{c}{Comments} \\ 
No. & (arcsec) & Flux  & source?  \\ 
\hline
 82 &  7.0 & 1.16 & y &  9\arcsec away from \chandra 28/\xmm 50.  \\
 91 & 12.5 & 0.94 & y &  Below final threshold. \\
 96 & 13.0 & 0.91 & y &  18\arcsec away from \chandra 77/\xmm 99. \\
102 & 13.0 & 0.80 & y &  Between \xmm 219 and \xmm 220. \\
    &      &      &   & (11\arcsec and 15\arcsec away respectively). \\
115 & 11.8 & 0.70 & y &  7\arcsec away from \chandra 107/\xmm 214 \\
121 & 10.4 & 0.66 & n &   \\
145 & 10.3 & 0.53 & y &  9\arcsec away from \chandra 61/\xmm 171. \\
149 &  7.1 & 0.50 & y &  15\arcsec away from \chandra 23/\xmm 39.\\
154 & 11.7 & 0.47 & n &  \\
158 & 11.5 & 0.46 & y &  13\arcsec away from \chandra 117/\xmm 78.\\
159 &  6.5 & 0.45 & y &  11\arcsec away from \chandra 20/\xmm 175.\\
163 & 12.1 & 0.43 & y &  \\
164 & 10.9 & 0.43 & n &  \\
171 &  8.6 & 0.37 & n &  \\
181 &  3.9 & 0.33 & n &  \\
182 & 11.6 & 0.32 & y &  Below final threshold. \\
183 &  7.4 & 0.31 & y &  Below final threshold. \\
185 &  7.0 & 0.31 & y &  \\
188 &  7.0 & 0.29 & y &  14\arcsec away from \chandra 52/\xmm43. \\
192 &  4.3 & 0.27 & y &  Below final threshold. \\
193 &  2.4 & 0.27 & y &  Below final threshold. \\
195 &  8.0 & 0.27 & n &  \\ 
197 &  7.5 & 0.26 & y &  Below final threshold. \\
198 &  5.5 & 0.26 & n &  \\
201 &  3.9 & 0.24 & n &   \\
203 &  3.8 & 0.23 & y &  12\arcsec away from \chandra 108/\xmm 103. \\
    &      &      &   &  In a region of extended emission.\\
204 &  7.4 & 0.23 & y &  18\arcsec away from \chandra 10/\xmm 119.  \\
205 &  3.3 & 0.23 & n &  \\
206 &  7.7 & 0.23 & y &  7\arcsec away from \chandra 142/\xmm 128.\\
207 &  7.8 & 0.22 & n &  \\
208 &  7.4 & 0.21 & n &  \\
210 &  1.0 & 0.19 & y &  Below final threshold. \\
211 &  9.5 & 0.19 & y  \\
213 &  7.2 & 0.19 & y & 40\arcsec away from \chandra 28/\xmm 50, \\
\hline 
\end{tabular}
\caption{List of \chandra sources with no \xmm counterpart. 
The \chandra flux is in units of $10^{-15}$ \ergs$\!\!$.
Excess emission associated with a
possible faint source is indicated by a `y' in column 4. 
The comment `Below final threshold' indicates a source which was detected with
EMLDETECT with DET\_ML$>5$ but subsequently
excluded from the final sourcelist according to the criteria in Section
\ref{sec:likes}. }
\label{tab:chandranoxmm}
\end{center}
\end{table*}

There are 70 \xmm sources with no \chandra counterpart. Of these, 9 sources
are either on the very edge or out of the field of view of the \chandra
mosaic \citep[see][]{mchardy03} and would therefore not be expected to 
have a \chandra counterpart. 
R-band images, centred on the positions of these 70 sources are presented in
Fig. \ref{fig:optimage}. (Optical images of those sources with a \chandra
counterpart are presented in \citealt{mchardy03}).
In 49 cases there is a clear optical counterpart or
counterparts within the positional error radius of the source.  Those sources
which have optical counterparts extended beyond 10\arcsec~ are shown in 
wider images covering $30\arcsec\times30\arcsec$ below the main image. Those in
the top panel have the same greyscale applied as in the main image, however in the
lower panel the greyscale has been individually adjusted for greater clarity.
Searching out
to a radius of 3\arcsec~ the number of sources with possible optical
counterparts is increased to 58. Of the 9 sources on the edge of, 
or outside, the \chandra field of view, 7 have an optical counterpart.

Table \ref{tab:xmmnochandra} lists the basic properties of
the 70 \xmm sources without \chandra counterparts.
The \chandra mosaic image was visually inspected at the positions of each 
\xmm source to see if any faint X-ray emission could be seen that was not 
formally detected in the \chandra source searching. Details of the number of 
counts observed within the \xmm positional error circle for each source are listed in column
10 and further comments are listed in column 11. Of the 61 sources not formally
detected within the \chandra field of view, faint emission is visible in 11
cases. None of the extended \xmm sources are detected with \chandra$\!\!$.
The \xmm observations are more sensitive to faint extended sources than the
\chandra observations.
One example is the \rosat source R117 \citep{mchardy98} 
which was detected with \xmm (number 56) but missed by \chandra$\!\!\!$. 
This source is a faint starburst galaxy with extended emission, probably on the
scale of the galaxy \citep{gunn01}. 

Conversely, there are 34 \chandra sources with no \xmm counterparts (within the
\xmm FOV). The properties of these sources are summarised in Table
\ref{tab:chandranoxmm}. Seven of these sources were originally detected in our initial
\xmm sourcelist, but removed from our final sourcelist after  we applied the
DET\_ML cutoffs derived in Section \ref{sec:likes}. Visual inspection of
the \xmm images at the remaining 27 \chandra source positions, suggests a 
faint source in 16 cases.
As well as the 4 close ($<10$\arcsec) pairs of sources found with \chandra
there are also  9 faint sources which were missed with \xmm because they are 
in the wings of much brighter sources.
This highlights the importance of our \chandra coverage in order to obtain accurate
source positions.

\section{Discussion}

\subsection{Source density in the \hr~ field}

Our \hr~ field represents one of the deepest blank field surveys with \xmm and
as such it is ideal for the investigation of faint X-ray source counts. 
Further, this work is the only \xmm based survey  published to date to
incorporate detailed Monte-Carlo modelling of the detection process allowing
accurate fitting of the source counts. Our source counts are inconsistent with
\emph{both} a single and double powerlaw fit in all energy bands, except for
our lowest energy band (0.2-0.5 keV) where a double powerlaw fit is
acceptable. However, a double powerlaw fit provides a 
better representation of the source counts according to the KS test 
in all but our hardest (5-10 keV) energy band where we are unable to 
constrain the break. 

Previous studies show a consensus in shape: the measured source counts are
described by a double powerlaw distribution flattening below a flux of
$\sim1\times10^{-14}$ \ergs and steeper than an Euclidean slope at fluxes
above the break ($S>S_{knee}$). However the measured normalisations differ by
up to $\sim30\%$ \citep{brandt01,rosati02}. \citet{yang03} 
attribute this field-to-field 
variation to the fact that
the fields studied cover small areas and are subject to cosmic
variance. Here the underlying clustered large scale structure imprints a
variation  in source density on small scales.

The slope of our soft band (0.5-2 keV) source counts, below the break
flux of $1.08 ^{+1.02}_{-0.39} \times10^{-14}$ \ergs$\!$, 
is consistent with recent
determinations using \chandra \citep{rosati02,bauer04,yang04}.
Our \hr~ field is the richest X-ray blank field observation reported to date 
in this energy band. 
Our overall normalisation is higher, though consistent with,
the CDF-N results of \citet{brandt01} and the HELLAS2XMM  counts of
\citet{baldi02}. However, the CDF-S source counts of \citet{rosati02},
and the observed counts in the Lockman hole \citep{hasinger98,hasinger01} 
are lower by $\sim$30\%. Allowing for the errors on the CDF-S source counts,
given by \citet{rosati02},
and Poisson errors on our counts, our field is still inconsistent with the
CDF-S at the 2.7$\sigma$ level. It is likely therefore that cosmic variance has
an important effect on the measured source counts found in deep X-ray
fields.

In the 2-5 keV energy band our observed counts are consistent with previous
\chandra and \xmm studies
\citep{brandt01,hasinger01,rosati02,baldi02,cowie02,yang04}. Our brightest source
is a factor of  $\sim10\times$ fainter than the limiting fluxes of the
serendipitous \asca surveys \citep{cagnoni98,ueda99,ueda99a} which makes
comparisons difficult. However, the \asca fits of \citet{cagnoni98} appear 
to be consistent with ours when extrapolated to the flux range covered by our
survey. Our overall 
normalisation at faint fluxes is closer to the CDF-S normalisation than it is to that
found in the  CDF-N in this energy
band. However, the CDF-S counts drop far more rapidly than ours towards
brighter fluxes.

In our hardest energy band (5-10 keV) the observed counts are consistent with a
Euclidean slope; $\gamma=2.80^{+0.67}_{-0.55}$. The slope agrees within the
errors with the results of \citet{cagnoni98}, \citet{fiore01}, \citet{baldi02} and the 
\xmm 5-10 keV counts in the Lockman hole \citep{hasinger01}.  
We expect a break in the source counts at a flux of $\sim4\times10^{-15}$ 
\ergs as reported from \chandra observations of the CDF-S \citep{rosati02}. 
It is therefore unsurprising that it is not detected in our survey, 
as we have no sources below this flux in our 5-10 keV sourcelist. 
At the brightest fluxes our counts are in agreement with findings from the
HELLAS2XMM survey \citep{baldi02}, \sax counts \citep{fiore01} and \asca 
counts \citep{cagnoni98}. 

\begin{figure}
\setlength{\unitlength}{1in}
\begin{picture}(3.5,3.5)
\put(0.0,0.2){\includegraphics{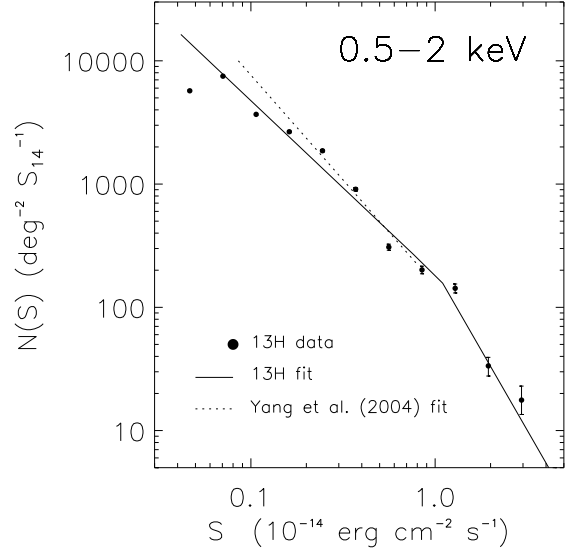}}
\end{picture}
\vspace{-1.0cm}
\caption{0.5-2 keV energy band differential \ns distribution. Overlaid is our
double powerlaw fit (solid line). Significant excesses over the model
distribution occur at a flux of $\sim2\times10^{-15}$ \ergs$\!$. Also shown is the
best fit model from the \chandra Survey of the Lockman
Hole-Northwest \citep{yang04}, which covers a similar flux range as our survey
(dotted line).}
\label{fig:diffcounts}
\end{figure}

\begin{figure}
\setlength{\unitlength}{1in}
\begin{picture}(3.5,3.5)
\put(0.0,0.2){\includegraphics{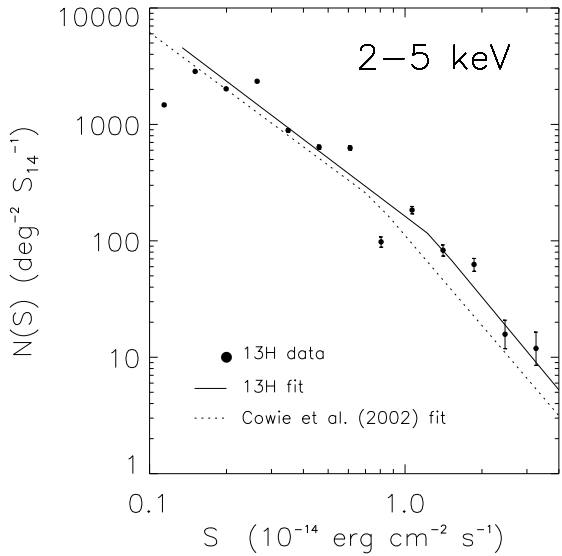}}
\end{picture}
\vspace{-1.0cm}
\caption{2-5 keV energy band differential \ns distribution. Our
best fit double powerlaw fit is overlaid (solid line). 
Significant excesses over the model distribution occur at a flux of
$\sim3\times10^{-15}$ and $\sim2\times10^{-14}$ \ergs$\!$. The best fit model of
\citet{cowie02} is also overlaid (dotted line) which is based upon fits to the
combined \chandra deep field counts and Hawaii survey fields SSA22 and
SSA13.}
\label{fig:diffcounts2}
\end{figure}

Although improved upon a single powerlaw, the double powerlaw fit to the source
counts is formally rejected in all but the softest energy band (0.2-0.5 keV). 
The differential counts show more clearly where
the deviations from the model occur and are
illustrated in Fig. \ref{fig:diffcounts} and Fig. \ref{fig:diffcounts2}. 
There is an
excess over the fit at a flux of $\sim2\times10^{-15}$ \ergs in both the 
0.5-2 keV and 2-5 keV energy bands. The fact that this excess is seen in both
energy bands suggests that the feature is most likely due to
clustering in the field rather than the X-ray spectral properties of the
sources around this flux. We defer a study of clustering to a later paper \citep{loaring04}.

We have also compared our source counts with wider area \chandra surveys which
we expect to  provide a more accurate 
value of the global source counts.
In the 0.5-2 keV energy band our fits to the
differential counts are similar to the fits obtained for the  
Lockman Hole-Northwest field \citep{yang04}. This implies
therefore that the CDF-S is a particularly underdense region. 
In the 2-5 keV energy band our double powerlaw fit lies above the fit of  
\citet{cowie02} which was obtained from the source counts in the CDF-S, CDF-N, 
and two Hawaii survey fields (SSA13 and SSA22) observed with \chandra$\!\!$. 
If the CDF-S data is excluded from their analysis their faint end 
normalisation agrees very well with ours, again suggesting that the CDF-S is
underdense.
This appears to contradict the earlier comparisons between the integral 2-5 keV
band counts in which our faint end normalisation closely matches the fits from
 \citet{rosati02}. However, the \citet{rosati02} CDF-S counts clearly drop below
ours at fluxes $> 5\times10^{-15}$ \ergs and it is at fluxes above the knee
that our counts are most discrepant from \citet{cowie02}. The comparisons
depend upon the adopted conversion between instrument energy bands and also the
range over which fitting is performed. \citet{rosati02} fit to much fainter
fluxes than \citet{cowie02}. As the source counts flatten at faint fluxes this
would result in an apparently higher normalisation at brighter fluxes
and may explain why the \citet{rosati02}
normalisation is higher than the \citet{cowie02} normalisation at fluxes around
$2\times10^{-15}$ \ergs$\!$. 
 
In converting values from the literature into our
energy bands we have used a photon index of $\Gamma=1.7$  which represents the
average slope of the XRB in the 0.2-12 keV energy range. 
Other authors have chosen to either use
individual photon indices measured from their sources  \citep{brandt01}
or a relatively flat $\Gamma=1.4$ \citep{rosati02} which represents the overall XRB spectral
slope. \citet{cowie02} use a photon index of $\Gamma=1.2$.
as they assume that the absorbed population must be significantly harder than
the unabsorbed population to produce the overall XRB spectrum. However, this
photon index
appears to be inappropriate at the fluxes probed by our survey. 
Considering the results from our X-ray
spectroscopy \citep{page03, page05}, the vast majority of sources have X-ray
spectra with a softer photon index than $\Gamma=1.2$, 
although they may be absorbed in the soft band.
In fact the value of the photon index used has little effect on source counts.
Using a photon index of 1.2 and 1.7 respectively to convert the 2-8 keV
counts of \citet{cowie02} into our 2-5 keV energy band results in a
\emph{integral} normalisation difference of $\sim$13\%  at our faintest fluxes
($2.2\times10^{-15}$ \ergs): counts derived assuming $\Gamma=1.7$ are higher
than those assuming $\Gamma=1.2$. This direct conversion does not account for
the fact that the original measured fluxes would have been derived 
from count rates using different photon indices. We have investigated this
effect further using spectra which were simulated in  XSPEC using the \chandra 
ACIS-I response matrix.  Assuming 
photon indices of $\Gamma=1.7$ and $\Gamma=1.2$ respectively to convert
count rates to fluxes,  we obtain  a flux
ratio $F_{1.7}/F_{1.2}$ of 0.9 in the 2-5 keV
energy band. Incorporating this factor into our flux
conversions reduces the apparent normalisation difference to $\sim$3\% at a
flux of  $2.2\times10^{-15}$ \ergs (with the normalisation of the $\Gamma=1.7$ 
source counts higher than that of the $\Gamma=1.2$ source counts). 
This factor is not large enough to account for the
measured differences in normalisation in the 2-5 keV energy band.
It is not necessary to employ any flux conversions for  the 0.5-2 keV energy
band. However, there is still a $\sim30\%$ discrepency in source counts between
the deep fields surveyed to date. One  concern is the cross-calibration between
the
different instruments used. However, \asca and \chandra fluxes agree at about
10\% and \chandra and \xmm fluxes agree at the  5\% level, and so the
differences in normalisation cannot be attributed to
instrument calibration offsets \citep{snowden01}. In the 0.5-2 keV band,   
we are confident that the field-to-field variations observed are real and due 
to cosmic variance. 

We find a larger field-to-field variation in the soft band (0.5-2 keV) 
source counts than in the hard band counts (2-5 keV). These findings are at
odds with the recent clustering measurements of \citet{yang03} who find
hard band sources to be more strongly clustered than soft band
sources. Further, \citet{gilli04} have found no
significant difference in clustering strength between soft and hard
sources. At present the evidence for any variation in clustering strength due
to X-ray hardness appears inconclusive. 
Clearly larger datasets are required to investigate this
issue further. Comparisons with wide area \xmm and \chandra serendipitous 
surveys coupled with further blank field observations should provide sufficient
signal to make significant advances in this area.

\subsection{Survey reliability and capabilities}
\label{sec:reliab}
\begin{table}
\begin{center}
\begin{tabular}{ccc}
\hline \hline 
 R magnitude & Expected number  & Observed number\\
\hline
18--19  & 0.5  &  10  \\
19--20  & 1.1  &  11  \\
20--21  & 2.4  &  18 \\
21--22  & 5.3  &  14 \\
22--23  & 11.7 &  61 \\
23--24  & 26.1 &  54 \\
24--25  & 57.7 &  42 \\
25--26  & 127.7 & 34 \\
\hline

\end{tabular}
\caption{The number of \xmm optical counterparts as a function of R
magnitude. X-ray -- optical cross matching was carried out within the 
95\% positional error circle of each \xmm
source and the number of optical detections 
are displayed as a function of R magnitude. 
Also shown is the number of expected chance
coincidences within each R magnitude interval 
given the optical source counts of \citet{mchardy03}.}
\label{tab:chance}
\end{center}
\end{table}

\begin{table}
\begin{center}
\begin{tabular}{ccc}
\hline \hline 
 Radio  Flux (mJy) & Expected number & Observed number \\
\hline
50--100 & $3.1\times10^{-3}$   & 1 \\
10--50  & $4.4\times10^{-2}$  & 0 \\
5--10  & $1.8\times10^{-2}$ & 0 \\
1--5  & 0.3 & 3 \\
0.5--1  & 0.1 & 4 \\
0.1--0.5  & 3.3 & 18 \\
0.05--0.10  & 2.0 & 6 \\
\hline
\end{tabular}
\caption{ The number of \xmm radio counterparts as a function of radio
flux. Radio -- optical cross matching was carried out within the 95\% positional error circle of each \xmm
source and the number of radio detections 
are displayed as a function of radio flux. 
Also shown is the number of expected chance coincidences given
the best-fit radio source count model of \citet{seymour04}.}
\label{tab:radio}
\end{center}
\end{table}

Our survey was specifically designed to study the X-ray source population
within one decade in flux either side of the break in the source counts. 
Contrary to recent deep \chandra  surveys
we aim not to resolve the entire XRB but rather to determine the dominant
emission mechanisms and amount of absorption in these faint sources. To
this end we require high quality X-ray spectra, in particular at hard 
energies to probe the absorbed population. The high throughput of \xmm at hard
energies makes it particularly suited to this kind of study. 

To efficiently study the faint source population we require a highly reliable
source catalogue with minimal spurious source contamination. We have examined
the quality of our final catalogue via extensive
Monte-Carlo simulations (Section \ref{sec:sims}).
Employing the likelihood cutoffs in each energy
band determined in Section \ref{sec:likes} we expect only 7 spurious sources 
to remain in our final source catalogue. Measurements of the fraction of flux 
amplified sources in our simulations indicate that confusion is unimportant  
($<2$\% of our sources are confused in all energy bands) 
at the fluxes probed by this survey. This is 
confirmed via our a cross-correlation with the \chandra catalogue of 
\citet{mchardy03}. We find that only 2\% of our sources have ambiguous 
\chandra counterparts.

The average input and output fluxes from our simulations at bright fluxes 
agree to within
1\%. This has important consequences for the determination of source counts,
and suggests any systematic deviation between input and output source counts
should be very small. 
Our simulations demonstrate that Eddington bias affects 
our measured source counts by at
most 23 per cent at the faintest fluxes. However, we correct for the effects of
Eddington bias when fitting our source counts, using the distribution of
$S_{out}/S_{inp}$ obtained from our simulations, as described in Section
\ref{sec:murdoch}.

Our simulations indicate that the offset in position 
between input and output sources is less than 2\arcsec~ in 68\% of cases within
the central 9\arcmin~ of the \xmm field of view (Section \ref{sec:simcomp}).
We therefore expect the majority of our sources to have positions accurate to 2\arcsec.
However, the EMLDETECT positional errors output from the detection chain are 
too small and an additional systematic positional error of 0.76\arcsec~ is required
(as determined in Section \ref{sec:comparison}) to match the
real and simulated positional error distributions. 
The high fidelity of our source positions and fluxes, coupled with
insignificant confusion illustrate the high reliability of the survey. 
We are able to construct useful X-ray spectra  down to a 2-5 keV band flux 
of $3\times10^{-15}$ \ergs$\!$, well beyond the initial survey goals \citep{page05}.

However, in order to extract the maximum scientific information from the survey, the
X-ray sources need to be optically identified, and compared with sources
detected in other wavebands, such as our radio source catalogue
\citep{seymour04}. This requires  highly accurate source
positions to identify the correct counterpart.
The additional systematic positional error has important consequences for
optical identification.  Using the optical 
source counts determined directly within the \hr~ field \citep{mchardy03}  
we have calculated the expected number of chance coincidences within an
appropriate radius of each source within the field (corresponding to the 95\%
positional error radius of each source).
Table \ref{tab:chance} shows the expected number of chance coincidences
for the 225 \xmm sources as a function of optical magnitude; it also shows the
actual number of optical sources observed within the error circles. 
X-ray/optical associations at
magnitudes $R\le23$ are firm, but approximately half of those at magnitudes
$23<R<24$ are likely to be spurious. At fainter magnitudes the
majority of optical counterparts are probably unrelated to the X-ray source.  
This highlights the
importance of the \chandra coverage over the field which provides 
accurate  X-ray positions and extends the range in which we can correctly select
an optical counterpart to a magnitude of $R=26$. This is an issue that must be 
considered in any deep \xmm survey since if ignored it may lead to a bias in
optical identifications. \chandra source counts determined by source type
find an increasing contribution of absorbed AGN and normal galaxies at the
fainter fluxes \citep{bauer04}. This suggests that
broad line AGN, which dominate at brighter fluxes, are most likely to be
correctly identified even without \chandra positions. 
However, in an \xmm survey of similar or greater depth than ours,
\chandra coverage is essential to identify correctly the optical counterparts
of the faintest sources, which are more likely to be associated with absorbed
AGN or normal galaxies. 

We have carried out a similar comparison with the \hr~ field radio catalogue of
\citet{seymour04}. The predicted number of chance coincidences (as predicted from the
best-fit starburst and AGN population model of \citealt{seymour04}) are
compared with the observed number of radio counterparts in Table \ref{tab:radio}.
We are able to securely identify the correct radio counterpart to a radio flux
of 0.1mJy without \chandra positions. Below this flux, we need \chandra
positions in order to reliably identify real associations between X-ray and radio
sources. The relative ease of the
identification of radio counterparts compared with that of optical counterparts
is due to the paucity of the radio source population.

Eddington bias most affects our counts at faint fluxes, where the statistical
errors on our fluxes are largest. For \xmm surveys with shorter exposure times,
source counts at faint fluxes will be significantly affected by Eddington bias.
Eddington bias at a particular flux depends on the relative errors on
the flux, and upon the slope and normalisation 
of the source counts distribution \citep{eddington1913,teerikorpi04}.
At a flux of $10^{-15}$ \ergs the relative difference in signal to noise 
between exposures of 120 and 40 ks is $\sim1.7$. 
Given that our source counts are boosted
by $\sim10$\% at $3\times10^{-15}$ \ergs we expect the source counts in a survey
with a 40 ks exposure time to be boosted by $\sim29$\% at $3\times10^{-15}$
\ergs$\!$. 
This is of particular relevance to shallower surveys such as the HELLAS2XMM
survey \citep{baldi02}, which should be heavily affected by Eddington bias at
faint fluxes.


\subsection{Contribution to the XRB}

\begin{figure}
\setlength{\unitlength}{1in}
\begin{picture}(3.5,3.5)
\put(0.0,0.5){\includegraphics{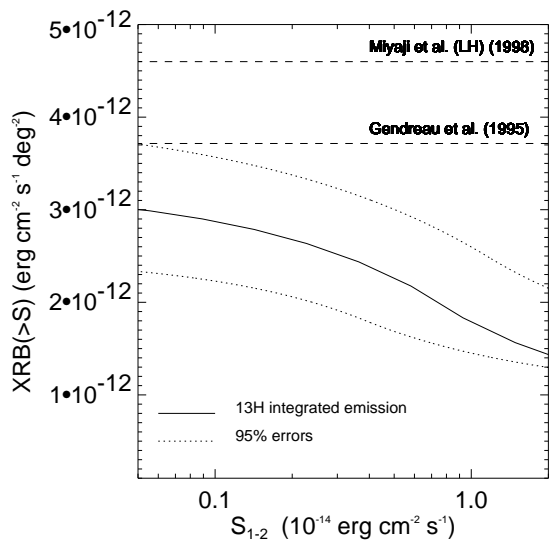}}
\end{picture}
\vspace{-1.0cm}
\caption{Discrete source contribution to the 1-2 keV X-ray background from
$0.1-1.0\times10^{-13}$ \ergs$\!\!$. The solid curve shows the integrated emission
from the best fit to our source counts, the dotted curves indicate the lower
and upper 95\% error on the integrated emission. Measurements of the  
XRB intensity taken from the literature are overlaid and labelled for
comparison. Our source counts account for $47-92$\% of the XRB intensity in
this energy band.}
\label{fig:xrblow}
\end{figure}

In this section we examine what fraction of the XRB we can probe with our
survey by comparing the integrated emission from our source counts with various
measurements of the XRB intensity.
To obtain the total 1-2 keV emission from point sources, we integrate our 0.5-2
keV differential source counts between fluxes of $2\times10^{-16}$ \ergs and $3\times 10^{-14}$ \ergs$\!$;
for fluxes greater than $3\times 10^{-14}$ \ergs we have added the
integrated counts obtained from the \rixos survey \citep{mason2000}. 
In the 1-2 keV band our source counts account for 51-100\% of the 
XRB intensity. Our contribution to the XRB as a function of flux is illustrated
in Fig. \ref{fig:xrblow}.
The main uncertainty in the resolved fraction lies in the range of reported
values for the absolute normalisation
of the XRB which are discrepant by up to 30\%. 
The XRB measurements of
\citet{gendreau95} and \citet{beppoxrb} are taken to represent the lower and
upper normalisations respectively in this energy band. 

\begin{figure}
\setlength{\unitlength}{1in}
\begin{picture}(3.5,3.5)
\put(0.0,0.5){\includegraphics{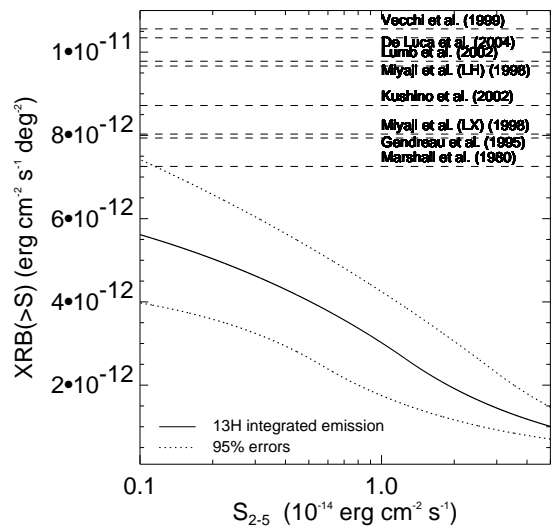}}
\end{picture}
\vspace{-1.0cm}
\caption{Discrete source contribution to the 2-5 keV X-ray background from
$0.1-1.0\times10^{-13}$ \ergs$\!\!$. The solid curve shows the integrated emission
from the best fit to our source counts, the dotted curves indicate the lower
and upper 95\% error on the integrated emission. Measurements of the  
XRB intensity taken from the literature are overlaid and labelled for
comparison. Our source counts account for $42-92$\% of the XRB intensity in
this energy band.}
\label{fig:xrb}
\end{figure}

In Fig. \ref{fig:xrb} we show  the
contribution of our sources to the integrated 2-5 keV XRB intensity. 
We integrated our differential source counts between fluxes 
of $2\times10^{-15}$ \ergs and $1\times 10^{-13}$
\ergs but this time we added on the integrated counts of \citet{cagnoni98}. 
With respect to the XRB measurements of \citet{gendreau95} and 
\citet{beppoxrb}, our source
counts contribute 50-93\% of the XRB intensity in the 2-5 keV band.
Our source counts would account for the whole 2-5 keV XRB intensity measured by
 \citet{marshall80}. Therefore, the XRB intensity within the \hr~ field must be
higher than that measured by  \citet{marshall80}. Indeed, given that our
source counts
are rich around $S_{knee}$, the XRB intensity within the \hr~ field may be
somewhat higher than the global average.

We have high quality X-ray spectra to 2-5 keV fluxes of $3\times10^{-15}$ \ergs
\citep{page03,page05}.
At these fluxes we have already resolved $\sim30$\% of the XRB, and hence the
\hr~ survey is readily suitable for the 
study of the nature and physical properties of the major XRB producing 
populations.

\section{Conclusions}
\label{conclusion}
In this paper we have presented the complete catalogue of 225 sources in
the \xmm \hr~ deep field which covers a sky area of 0.18 deg$^{2}$ centred on
position 13h 34m +37$^\circ$ 53'. 
We reach source densities of 700
$\rm{deg^{-2}}$ at a flux of $4.1\times10^{-16}$ \ergs in our 
lowest energy band. 
In the 0.5-2 keV band we find source densities of 1300 at a limiting flux of
$4.5\times10^{-16}$ \ergs$\!\!$. At harder energies we reach fluxes a factor
$\sim10\times$ brighter, with  source densities of 900 and 300 $\rm{deg^{-2}}$
at limiting fluxes of $1.1\times10^{-15}$ \ergs and  $5.3\times10^{-15}$ \ergs
in the 2-5 keV and 5-10 keV energy bands respectively.

We have carried out extensive simulations of the detection process in order to
assess the reliability of our source catalogue. 
Simulations indicate that confusion is small ($<2$\%) in our images. 
We have curtailed the
sourcelist directly derived from the SAS  according to the
optimal statistical detection likelihoods in each band determined from our
simulations. We expect $<7$ spurious  sources to
remain in the final catalogue.

Within the central 9\arcmin~ of the \xmm field of view, positional errors
are less than 2\arcsec~ for 68\% of our simulated sources and our
input/output fluxes agree to within 1\% at bright fluxes. 
Comparison of the input/output source positional offsets from our simulations
with the positional offsets found between the \xmm and \chandra counterparts
suggest that an additional systematic error of 0.76\arcsec~ should be added in
quadrature to
the EMLDETECT positional errors.  This poses no problem for radio counterpart
identification. We match our catalogue with the radio catalogue of
\citet{seymour04} and are confident in our choice of radio counterpart 
to a flux of 0.1mJy. In the optical we are 
confident in our choice of optical counterparts to a magnitude of $R=23$ using
our \xmm positions. For magnitudes fainter than this we need \chandra positions
which allow reliable identification of optical counterparts to $R=26$
\citep{mchardy03}. 
 
We have computed the best-fit parameters for the differential  \ns function
using a method adapted from \citet{murdoch73}. 
In all but our hardest energy band the data
are described by a double powerlaw model with a break flux at $\sim 10^{-14}$
\ergs$\!\!$. The counts below the break are flatter than the Euclidean
case. 
Our measured source counts are in broad agreement with previous
determinations from the \chandra deep fields \citep{brandt01,rosati02}, \xmm
\citep{hasinger01, baldi02} and \sax \citep{fiore01} surveys. The
overall normalisation in the 0.5-2 keV band is similar to, though higher than,
that found in  the CDF-N survey \citep{brandt01}. 
Both the  CDF-S \citep{rosati02} and Lockman hole \citep{hasinger01} 
normalisations are significantly lower than found here. This field-to-field
variation in source density may be attributed to the underlying clustered large
scale structure which imprints a variation in source density on small scales. 

In the 2-5 keV band our faint end normalisation is consistent with the
CDF-N \citep{brandt01}, CDF-S \citep{rosati02} and \xmm Lockman hole counts
\citep{hasinger01}. There is minimal flux overlap with the  \asca surveys 
\citep{cagnoni98,ueda99,ueda99a}. The 2-5 keV band \asca counts lie above 
those found in this survey \citep{cagnoni98} although they are still 
consistent with our model fits.  
In our hardest energy band (5-10 keV) our counts are in broad agreement with
those from \sax and \asca, and with the  5-10 keV counts in the Lockman hole
\citep{hasinger01}. Again there is little overlap with the brighter \asca and
\sax surveys.

The sources in our survey straddle the break in the source counts. 
Accounting for the uncertainty in our source counts and the absolute
normalisation of the XRB we resolve 51-100(50-93)\% of the 1-2 (2-5) keV XRB
emission. At the break in the source counts we resolve $\sim30$\% of the 2-5
keV XRB. At these fluxes we have X-ray spectra and are therefore able
to study the emission mechanisms and investigate absorption
in a  significant fraction of the X-ray source population.

\onecolumn
\begin{landscape}
\small
\setlength{\tabcolsep}{0.5mm}

\begin{longtable}[l]{r@{\hspace{0.2cm}}r@{\hspace{0.2cm}}r@{\hspace{0.2cm}}r@{\hspace{0.2cm}}r@{\hspace{0.2cm}}r@{\hspace{0.2cm}}r@{\hspace{0.2cm}}r@{\hspace{0.2cm}}r@{\hspace{0.2cm}}r@{\hspace{0.2cm}}r@{\hspace{0.2cm}}r@{\hspace{0.2cm}}r@{\hspace{0.2cm}}r@{\hspace{0.2cm}}r@{\hspace{0.2cm}}r@{\hspace{0.2cm}}r@{\hspace{0.2cm}}r@{\hspace{0.2cm}}r@{\hspace{0.2cm}}r@{\hspace{0.2cm}}r@{\hspace{0.2cm}}r@{\hspace{0.2cm}}r@{\hspace{0.2cm}}r@{\hspace{0.2cm}}r}



\caption{\hr~ \xmm deep field catalogue. All fluxes are in units of
$10^{-15}$ \ergs$\!\!$. The columns labelled `Counts' and `Counts error' 
give the best fit number of source counts and the 1$\sigma$ error as output 
from EMLDETECT. Flux errors are at 1$\sigma$ confidence.
If in any energy band there are fewer counts at the source position 
in the image than in the background map, then the number of
source counts and the source flux are recorded as zero for that band.
In this case the `Counts error' and 'Flux error' columns represent the
1$\sigma$ upper limit. The `Pos error' column gives the EMLDETECT positional
error (1$\sigma$) in arc seconds. The `Offaxis angle' column gives the  \xmm
offaxis angle in arc minutes. The `\chandra No.' column gives the
\chandra catalogue number taken from
\citet{mchardy03}. The `\chandra offset' column gives the offset in position 
between the \xmm and \chandra source, in arcsec.}
\\
\hline 
\hline 

 &  & & & &  \multicolumn{2}{c}{0.2-0.5 keV}
&  \multicolumn{2}{c}{0.5-2.0 keV} &  \multicolumn{2}{c}{2.0-5.0 keV}
&  \multicolumn{2}{c}{5.0-10.0 keV} &   \multicolumn{2}{c}{0.2-0.5 keV} 
&  \multicolumn{2}{c}{0.5-2.0 keV} &
\multicolumn{2}{c}{2.0-5.0 keV} &  \multicolumn{2}{c}{5.0-10.0 keV}
&\multicolumn{2}{c}{0.2-10.0 keV}   \\
\multicolumn{3}{c}{\xmm} &  
\multicolumn{1}{c}{Pos$^{a}$} &
\multicolumn{1}{c}{Offaxis$^{b}$} & &
\multicolumn{1}{c}{Counts} & &
\multicolumn{1}{c}{Counts} & &
\multicolumn{1}{c}{Counts} & &
\multicolumn{1}{c}{Counts} & &
\multicolumn{1}{c}{Flux} & & 
\multicolumn{1}{c}{Flux} & & 
\multicolumn{1}{c}{Flux} & & 
\multicolumn{1}{c}{Flux} & & 
\multicolumn{1}{c}{Flux} & \multicolumn{2}{c}{\chandra} \\

\multicolumn{1}{c}{No.} 
& \multicolumn{1}{c}{Ra J(2000)} 
& \multicolumn{1}{c}{Dec J(2000)}   
& \multicolumn{1}{c}{error} & \multicolumn{1}{c}{angle} 
& \multicolumn{1}{c}{Counts} & \multicolumn{1}{c}{error} 
& \multicolumn{1}{c}{Counts}  & \multicolumn{1}{c}{error}
& \multicolumn{1}{c}{Counts} & \multicolumn{1}{c}{error} 
& \multicolumn{1}{c}{Counts}  & \multicolumn{1}{c}{error} 
& \multicolumn{1}{c}{Flux}  & \multicolumn{1}{c}{error} 
& \multicolumn{1}{c}{Flux}  & \multicolumn{1}{c}{error} 
& \multicolumn{1}{c}{Flux}  & \multicolumn{1}{c}{error}
& \multicolumn{1}{c}{Flux}  & \multicolumn{1}{c}{error} 
& \multicolumn{1}{c}{Flux}  & \multicolumn{1}{c}{error} 
& \multicolumn{1}{c}{No.$^{c}$} & \multicolumn{1}{c}{offset$^{d}$} \\
\hline
\endfirsthead
\hline \hline  
&  & & & &  \multicolumn{2}{c}{0.2-0.5 keV}
&  \multicolumn{2}{c}{0.5-2.0 keV} &  \multicolumn{2}{c}{2.0-5.0 keV}
&  \multicolumn{2}{c}{5.0-10.0 keV}    & \multicolumn{2}{c}{0.2-0.5 keV} 
&  \multicolumn{2}{c}{0.5-2.0 keV} &
\multicolumn{2}{c}{2.0-5.0 keV} &  \multicolumn{2}{c}{5.0-10.0 keV}
&\multicolumn{2}{c}{0.2-10.0 keV}   \\

\multicolumn{3}{c}{\xmm} &  
\multicolumn{1}{c}{Pos$^{a}$} &
\multicolumn{1}{c}{Offaxis$^{b}$} & &
\multicolumn{1}{c}{Counts} & &
\multicolumn{1}{c}{Counts} & &
\multicolumn{1}{c}{Counts} & &
\multicolumn{1}{c}{Counts} & &
\multicolumn{1}{c}{Flux} & & 
\multicolumn{1}{c}{Flux} & & 
\multicolumn{1}{c}{Flux} & & 
\multicolumn{1}{c}{Flux} & & 
\multicolumn{1}{c}{Flux} & \multicolumn{2}{c}{\chandra} \\

\multicolumn{1}{c}{No.} 
& \multicolumn{1}{c}{Ra J(2000)} 
& \multicolumn{1}{c}{Dec J(2000)}    
& \multicolumn{1}{c}{error} & \multicolumn{1}{c}{angle} 
& \multicolumn{1}{c}{Counts} & \multicolumn{1}{c}{error} 
& \multicolumn{1}{c}{Counts}  & \multicolumn{1}{c}{error}
& \multicolumn{1}{c}{Counts} & \multicolumn{1}{c}{error} 
& \multicolumn{1}{c}{Counts}  & \multicolumn{1}{c}{error} 
& \multicolumn{1}{c}{Flux}  & \multicolumn{1}{c}{error} 
& \multicolumn{1}{c}{Flux}  & \multicolumn{1}{c}{error} 
& \multicolumn{1}{c}{Flux}  & \multicolumn{1}{c}{error}
& \multicolumn{1}{c}{Flux}  & \multicolumn{1}{c}{error} 
& \multicolumn{1}{c}{Flux}  & \multicolumn{1}{c}{error} 
& \multicolumn{1}{c}{No.$^{c}$} 
& \multicolumn{1}{c}{offset$^{d}$} \\
\hline
\endhead
\hline & & & & & \multicolumn{3}{r}{{Continued on next page}} \\ 
\endfoot
\hline
\endlastfoot

  1  & 13 33 28.96&37 55 58.82&  1.57& 12.91&   20.08&    5.87&   55.01&    9.56&   13.04&    6.11&    1.36&    3.16&   3.39&   0.99&   6.80&   1.18&   2.89&   1.35&   1.78&   4.12&  14.86&  4.60\\
  2  & 13 33 30.32&37 56 12.21&  2.38& 12.67&   14.12&    5.70&   22.65&    6.82&    8.47&    5.14&    0.00&    5.05&   1.93&   0.78&   2.37&   0.71&   1.65&   1.00&   0.00&   5.44&   5.94&  5.63\\
  3  & 13 33 30.68&37 55 05.57&  3.04& 12.53&    2.06&    3.29&   23.89&    7.07&    0.00&    3.95&    0.00&    2.40&   0.25&   0.40&   2.30&   0.68&   0.00&   0.72&   0.00&   2.42&   2.55&  2.64\\
  4  & 13 33 32.24&37 58 15.92&  1.33& 12.67&    0.00&    2.20&   42.24&    8.16&   52.13&    9.52&    3.62&    5.30&   0.00&   0.16&   2.60&   0.50&   6.77&   1.24&   2.32&   3.39&  11.68&  3.65& 47& 1.29\\
  5  & 13 33 34.66&37 49 10.68&  1.27& 13.08&   22.60&    5.06&   49.76&    8.22&   19.01&    6.10&    1.49&    3.24&   2.61&   0.58&   5.54&   0.92&   5.36&   1.72&   1.67&   3.63&  15.18&  4.17& 49& 2.70\\
\end{longtable}
\end{landscape}
\twocolumn

\begin{appendix}

\section{Synthetic field generation}

\subsection{Method}

Input sourcelists were generated independently for each of the four energy
bands. The differential source count function, \ns, was modelled as a broken
power law distribution: 

\begin{equation}
N(S) = \left\{ \begin{array}{lr}
                      K_1 \left( \frac{S}{S_{norm}} \right) ^{-\gamma _1} & \quad ( S<S_{knee}) \\
                      \quad & \quad \\
                      K_2 \left( \frac{S}{S_{norm}} \right) ^{-\gamma _2} & \quad (S\ge S_{knee}) 
                \end{array}
        \right.
\label{equation_NofS}
\end{equation}

The normalisation constants were chosen such that the 
function is continuous at the knee, i.e. $K_1/K_2 = (S_{knee}/S_{norm})^{\gamma
_1 - \gamma _2}$.
The parameters selected for the simulations were based on results from the
CDF-S \citep[see][]{rosati02}. A photon index, $\Gamma = 1.7$ (approximately 
that of the 
XRB in the 0.2-10 keV energy range), was used to convert from the \chandra
energy bands to the energy bands used in this study. 
We have slightly modified the \chandra \ns parameters via an iterative process
so that the `output' source counts approximately match those seen in the 
\hr~ field data.

In order to incorporate the effects of source confusion, we set our simulated
source flux limits, $S_{lim}$, to values approximately five times fainter than
the limit reached by the \hr~ data.
The mean number of input sources per field per energy band,
$\langle N \rangle$, was  calculated by integrating \ns from
$S_{lim}$ to infinity.
The actual number of sources used in each simulated field was taken 
randomly from a Poisson distribution about $\langle N \rangle$. 
For each of these sources, a flux was randomly assigned from the 
appropriate $N(S)$ distribution. 
Any input source having a flux greater than twice that of the brightest source
in the \hr~ data was discarded.
This prevented any single simulated field being dominated by an extremely
bright source: a situation not seen in the \hr~ data.
The effects of source clustering are ignored in this analysis, 
hence each input source was assigned a purely random position within the field.
Finally, the source fluxes were converted to count rates.

\subsection{Imaging characteristics}
To convert the simulated input sourcelists to images, 
one has to take account of the complex point spread function (PSF) of the 
EPIC cameras.
We used the `MEDIUM' accuracy PSF description \citep{kirch04} from the
\xmm calibration files,
which is also the model used by the SAS source searching task EMLDETECT.
This PSF description consists of a number of small ray-traced maps each 
describing
the PSF at a particular energy and off-axis angle, covering the full EPIC field
of view and energy range. 
These maps were interpolated in energy and off-axis angle, allowing us to
evaluate the fraction of any source's flux within any image pixel.  
The SAS-generated exposure maps from the \hr~ field give the effective
exposure times and vignetting corrections for each energy band and EPIC camera.
The simulated images were generated pixel by pixel by summing the contribution
from all input sources.
We added a two-component, (vignetted and un-vignetted), synthetic background to
the simulated images to reproduce that observed in the \hr~ field.
The correct level of this background was determined through an iterative
process because undetected, faint simulated sources contribute significantly to the
diffuse background.
Finally, the value of each pixel was randomly drawn from the appropriate
Poisson distribution to simulate photon counting noise.

\subsection{Simulated image analysis}
We used a single-band version of the source searching
process (as described in Section \ref{sec:background}) on the simulated images.
This process was repeated independently in each of the four energy bands.
Two iterations of the background fitting/source finding process were carried
out per band per field. The output sourcelist was curtailed at a detection
likelihood of DET\_ML = 5.

\section{Assessing the impact of confusion}

\begin{figure}
\setlength{\unitlength}{1in}
\begin{picture}(3.5,3.5)
\put(-0.5,-2.0){\includegraphics{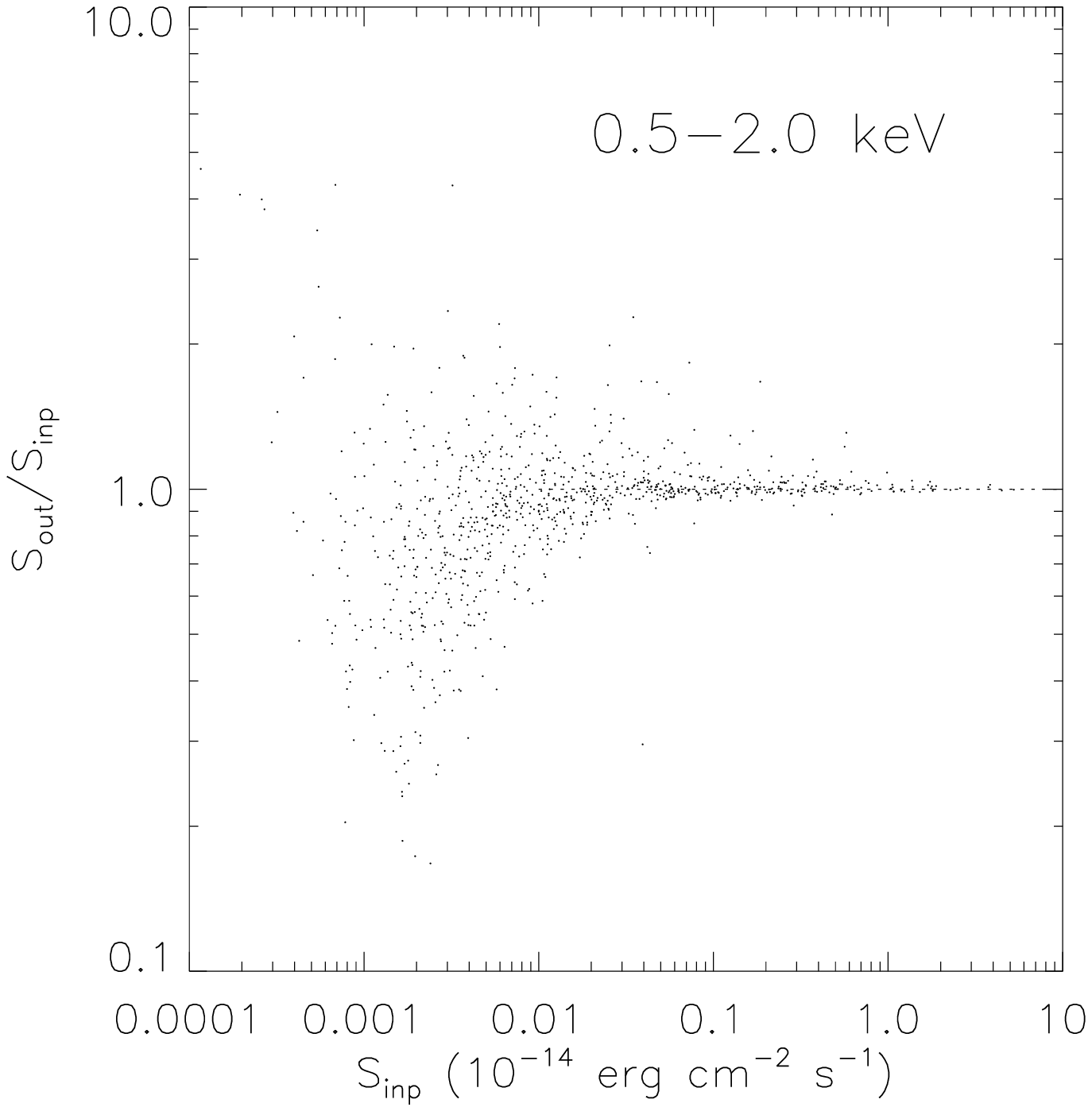}}
\end{picture}
\vspace{-1.0cm}
\caption{Scatter plot of input vs output flux 
in the 0.5-2 keV energy band. We only show those
detected sources having $\rm{DET\_ML \ge 5}$ and valid input counterparts within
5\arcsec, 8\arcsec, and 10\arcsec~ for input offaxis angles of $0-9'$, $9-12'$
and $>12'$ respectively.}
\label{fig:fluxinoutfaint}
\end{figure}

\begin{figure}
\setlength{\unitlength}{1in}
\begin{picture}(3.5,3.5)
\put(0.0,0.5){\includegraphics{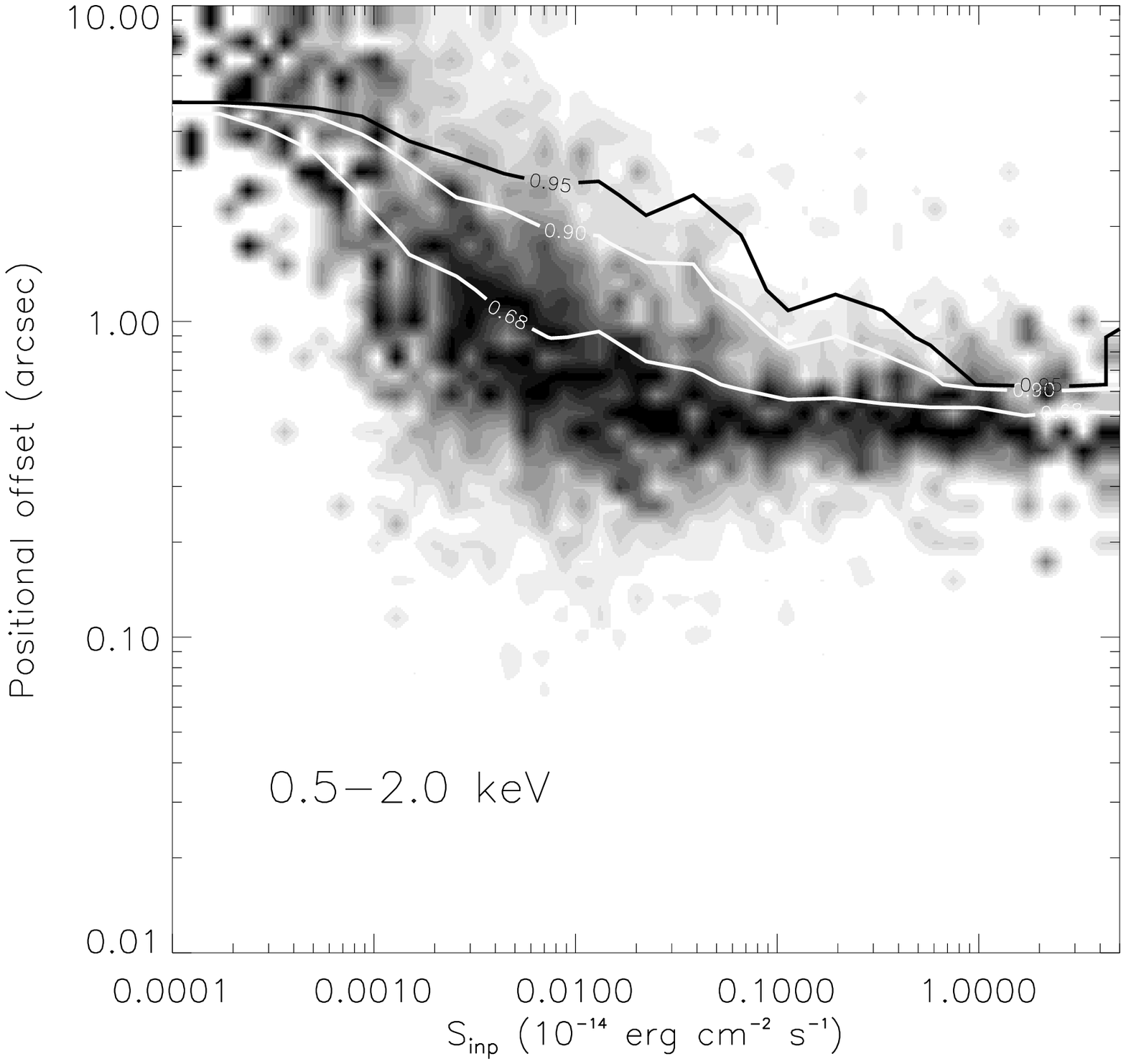}}
\end{picture}
\vspace{-1.0cm}
\caption{Greyscale image showing the distribution of positional offsets
between output and input source locations as a function of input flux in the 
0.5-2 keV energy band. 
Those sources within the central
9\arcmin~ of the \xmm field of view having $ \rm {DET\_ML \geq
5 }$ are shown. The concentration of positional offsets as a function of
$S_{inp}$ is indicated by the darkness of the greyscale image.
The three contours show the distances within which
68, 90, and 95 per of the data lie.}
\label{fig:contourfaint}
\end{figure}

\subsection{Simulation method}
In order to investigate the ultimate confusion limit of deep \xmm surveys,
we have produced a small number of extremely deep simulations, with exposure
times 1000$\times$ that of the \hr~ field. The input sourcelists reach
fluxes of a few $10^{-19}-10^{-18}$ \ergs depending on energy band. 
No background or Poisson noise were added to the
images in these simulations, so differences between the input and output source
counts  arise solely as a result of confusion. The sources were matched within
a cutoff radius, $r_{cut}$, which depends on \xmm offaxis angle as described in 
Section \ref{sec:simcomp}.

\subsection{Results}

Fig. \ref{fig:fluxinoutfaint} shows the distribution of $S_{out}/S_{inp}$ for
our  10 simulations in the 0.5-2 keV energy band. The scatter on $S_{out}/S_{inp}$ increases
dramatically below a flux of $10^{-16}$ \ergs.  At similar fluxes, the input-output position
offsets, shown in Fig. \ref{fig:contourfaint} become much larger. The source
density at this flux level is approximately 2000 deg$^{-2}$ and represents the
limit beyond which source properties cannot be recovered reliably irrespective
of exposure time.

\begin{figure*}
\setlength{\unitlength}{1in}
\begin{picture}(7,5.5)
\put(-0.2,5.5){\includegraphics{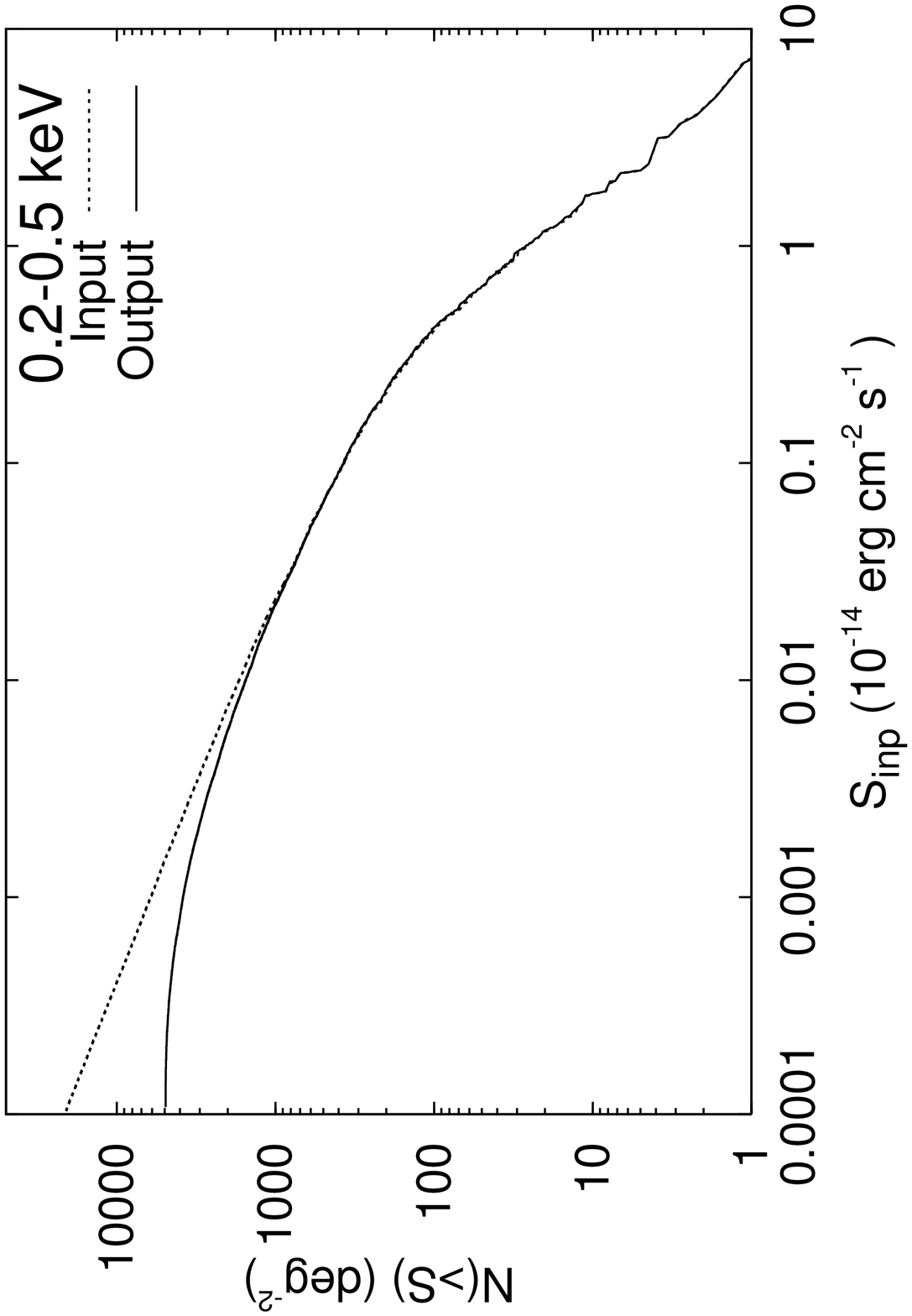}}
\put(3.2,5.5){\includegraphics{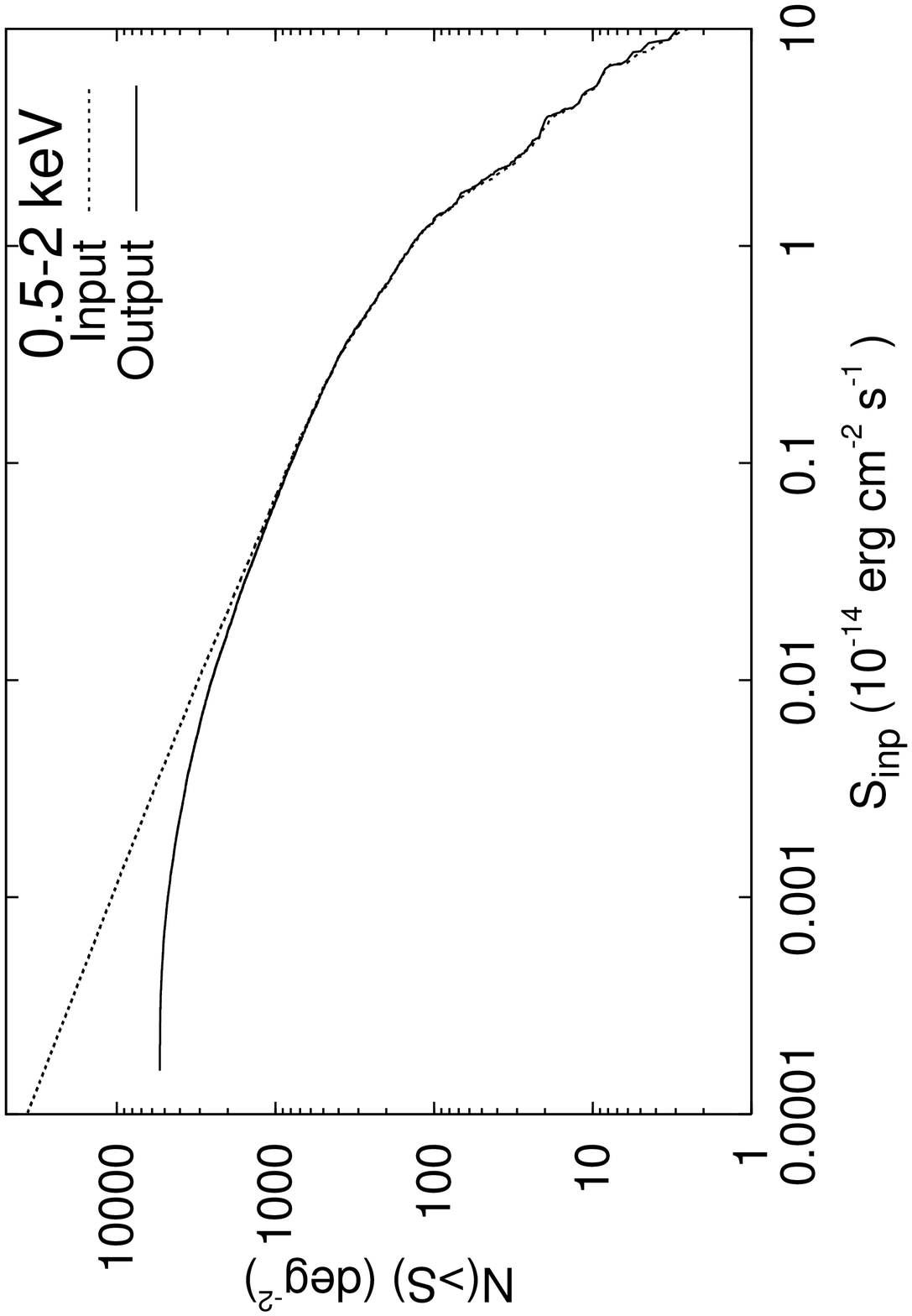}}
\put(-0.2,3.0){\includegraphics{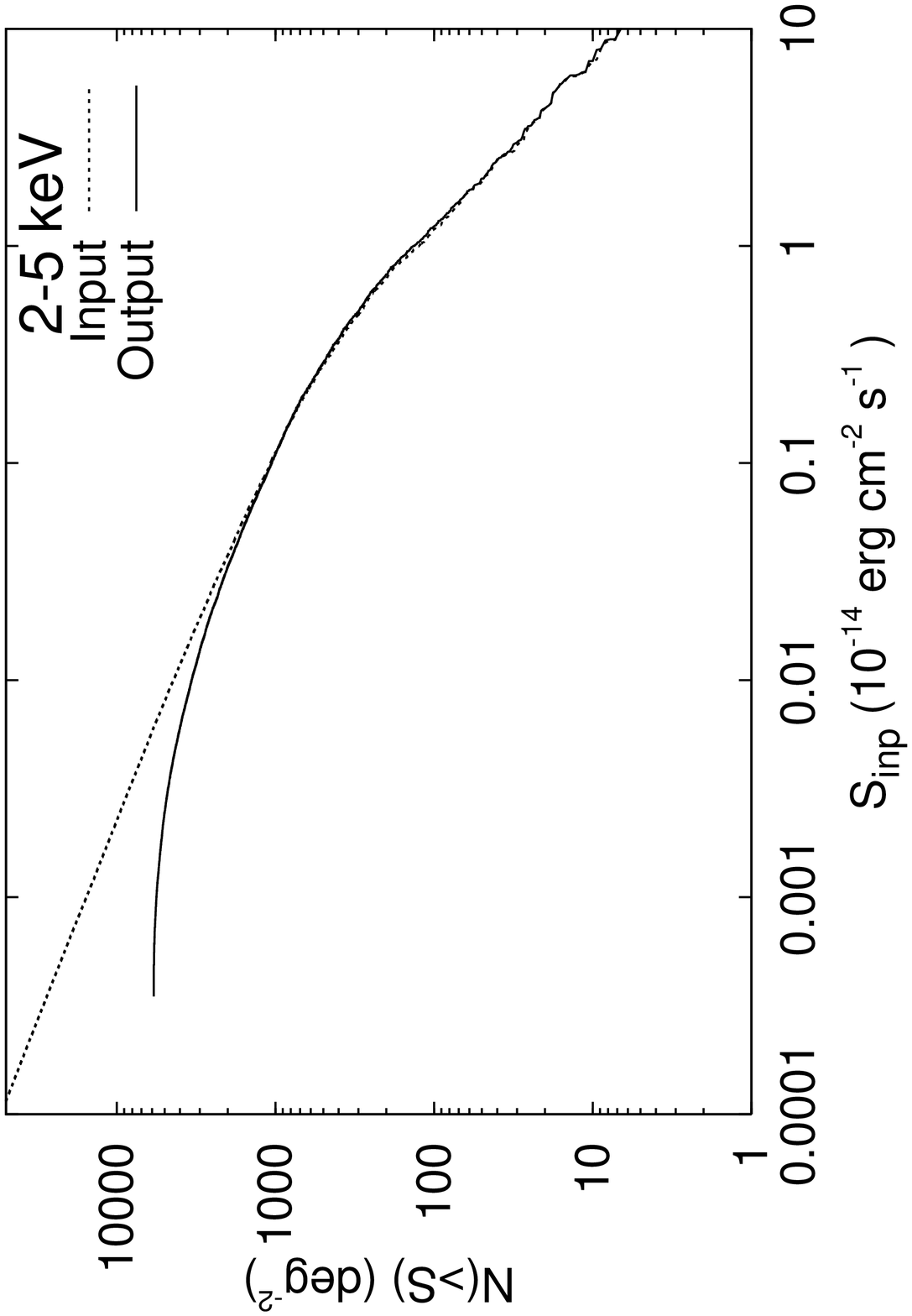}}
\put(3.2,3.0){\includegraphics{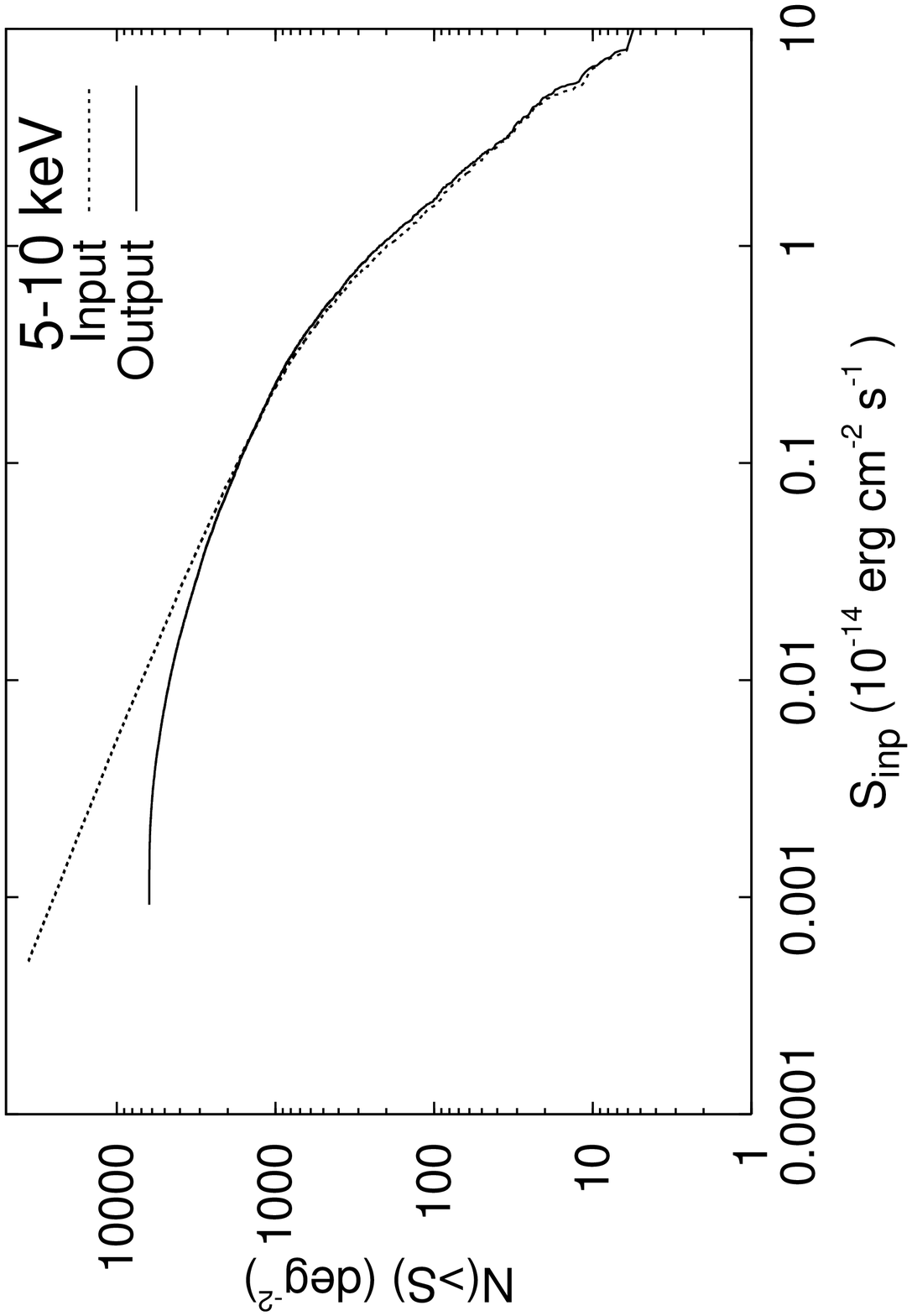}}

\end{picture}
\vspace{-1.0cm}
\caption{Comparison of input (dotted line) and output (solid line) simulated
source counts. Input source counts were simulated to fluxes of
$5\times10^{-19}$ \ergs in the 0.2-0.5 keV and 0.5-2 keV energy bands. In the
2-5 keV and 5-10 keV energy bands the input counts were simulated to fluxes of 
$1\times10^{-18}$ and $5\times10^{-18}$ \ergs. 
Confusion is evident at faint fluxes where the input and output source
counts rapidly diverge.}

\label{fig:deepns1}
\end{figure*}

Fig. \ref{fig:deepns1} show the average input and output integral source
counts in each of our four energy bands. 
The input and output source counts are already discrepant
 when the source density reaches 2000 deg$^{-2}$, and diverge rapidly at
fainter fluxes. The approximate input fluxes corresponding to source densities
of 2000 deg$^{-2}$
are $3\times10^{-17}$, $10^{-16}$, $2\times10^{-16}$ and 
$3\times10^{-16}$ \ergs in the 0.2-0.5, 0.5-2, 2-5 and 5-10 keV energy
bands respectively.

\end{appendix}

\end{document}